\documentclass{article}

\usepackage{hyperref}
\usepackage{amsmath,amsfonts,amssymb}
\usepackage{graphicx}
\usepackage{natbib}
\usepackage[margin=0.9in]{geometry}
\providecommand{\keywords}[1]{\textbf{\textit{Key Words---}} #1}

\usepackage{fancyhdr}
\graphicspath{{pdf/}}

\newcommand{\eqnref}[1]{Equation (\ref{#1})}
\newcommand{\figref}[1]{Figure \ref{#1}}
\newcommand{\eqnsref}[1]{Equations (\ref{#1})}

\def\bx{{\bf x}}
\def\bp{{\bf p}}
\def\bpa{{\bf p}_A}
\def\bpad{{\bf p}_A'}

\def\bxa{{\bf x}_A}
\def\bxb{{\bf x}_B}
\def\bxs{{\bf x}_S}
\def\bxr{{\bf x}_R}
\def\bpa{{\bf p}_A}

\def\bxjs{{\bf x}_{S,j}}

\def\d3{\partial_3}
\def\dt{\partial_t}

\begin{document}
\title{3D Marchenko applications: Implementation and examples}
\author{Joeri Brackenhoff\footnote{Corresponding author, email: johannes.brackenhoff@erdw.ethz.ch} \footnote{Delft University of Technology, Department of Geoscience and Engineering, Stevinweg 1, 2628CN Delft, the Netherlands} \footnote{ETH Z\"urich, Seismology and Wave Physics group,  Sonneggstrasse 5, 8092 Zurich, Switzerland}, Jan Thorbecke\footnotemark[1], Giovanni Meles\footnotemark[1] \footnote{University of Lausanne, Institute of Earth Sciences, Lausanne, 1015, Switzerland},\\ Victor Koehne\footnote{Senai Cimatec, Av. Orlando Gomes 2845, Salvador, Bahia, 41650-010, Brazil}, Diego Barrera\footnotemark[5]\hspace{5pt} and Kees Wapenaar\footnotemark[1]}

\maketitle

\section*{Acknowledgments}
The authors wish to thank Jonathas Maciel, Reynam Pestana, Ot\'avio Ribeiro, Adhvan Novais Furtado and Jo\~ao Marcelo Souza for their contributions to this work, as well as the SENAI CIMATEC Supercomputing Center for their support of this research. Additionally, we would also like to thank associate editor Matteo Ravasi and two anonymous reviewers for their constructive comments and feedback.\\
This work has received funding from the European Union's Horizon 2020 research and innovation program: European Research Council (grant agreement no. 742703).

\section*{Conflicts of interest}
The authors declare no conflicts of interest.

\begin{abstract}
We implement the 3D Marchenko equations to retrieve responses to virtual sources inside the subsurface. For this, we require reflection data at the surface of the Earth that contain no free-surface multiples and are densely sampled in space. The required 3D reflection data volume is very large and solving the Marchenko equations requires a significant amount of computational cost. To limit the cost, we apply floating point compression to the reflection data to reduce their volume and the loading time from disk. We apply the Marchenko implementation to numerical reflection data to retrieve accurate Green's functions inside the medium and use these reflection data to apply imaging. This requires the simulation of many virtual source points, which we circumvent by using virtual plane-wave sources instead of virtual point sources. Through this method, we retrieve the angle-dependent response of a source from a depth level rather than of a point. We use these responses to obtain angle-dependent structural images of the subsurface, free of contamination from wrongly imaged internal multiples. These images have less lateral resolution than those obtained using virtual point sources, but are more efficiently retrieved.
\end{abstract}
\keywords{Signal processing, Seismics, Numerical study}

\section*{Data Availability}
The SEG/EAGE overthrust model and additional information can be downloaded from \url{https://wiki.seg.org/wiki/SEG/EAGE_Salt_and_Overthrust_Models}. The model is licensed under the Creative Commons Attribution 4.0 International License. To view a copy of the license, visit https://creativecommons.org/licenses/by/4.0 or send a letter to Creative Commons, PO Box 1866, Mountain View, CA 94042, USA. The license permits any user to freely copy and redistribute the material in any medium or format. Users are free to remix, transform, and build upon the material for any purpose, including commercially. \\
The software for the finite-difference modeling, the Marchenko method and the Eikonal solver can all be found at \url{https://github.com/JanThorbecke/OpenSource}. Demo scripts for the methods in this paper can be found in the directory \texttt{marchenko3D/demo/marchenko3D/oneD/} in the OpenSource directory. The 3D figures in this paper were created using the ParaView software \cite{Ayachit2015}. The data that are used in this paper are generally stored in the format of Seismic Unix (SU) by \cite{SU2016}.

\section*{Introduction}
The Marchenko method, originally derived for the field of quantum physics \citep{marchenko1955reconstruction}, was introduced in the field of geophysics about a decade ago. The method employs reflection data that are recorded at the surface of the Earth to create sources and receivers inside the Earth while internal multiples are handled correctly. Because these sources and receivers are not physically located inside the subsurface but instead are created from the reflection data, they are called virtual sources and receivers. The original Marchenko equation was introduced in the field of quantum physics in the 1950s by \cite{marchenko1955reconstruction}. More recently, the method was developed in the field of geophysics by \cite{broggini2012connection} for 1D media and was further extended for 2D and 3D applications by \cite{wapenaar2013three}, \cite{Broggini2014}, \cite{slob2014seismic} and \cite{Behura2014}. These authors showed that the reflection data and the Green's function, i.e. the impulse response of a medium, can be related via a so-called focusing function, which is a wavefield that focuses from an open boundary to a location in the subsurface, often referred to as the focal location. This relation can be rewritten into a Marchenko-type equation in 3D and subsequently be solved using only the reflection data and an estimation of the first arrival of a wavefield from the focal location. The first arrival can be easily obtained through the use of a background velocity model.

The Marchenko method has been used for many applications since it was introduced in the field of geophysics. It has been applied to create images of the subsurface that are free of artifacts related to the internal multiples \citep{Broggini2014,meles2015internal,Ravasi2017,Matias2018}. However, the method has also been used to obtain the homogeneous Green's function in the subsurface \citep{brackenhoff2019,brackenhoff2019virtual}, remove internal multiples from reflection data \citep{ZhangStaring2018} or to retrieve plane-wave responses in the subsurface \citep{Meles2018}. These applications made use of acoustic reflection data that contained no free-surface multiples. It has been shown that the Marchenko method can also be applied using elastic reflection data \citep{costa2014elastodynamic,wapenaar2014single, zhou2019,reinicke2020elastodynamic} and reflection data that contain free-surface multiples \citep{singh2015marchenko,SlobEAGE2017}. Although the theory has been fully developed for 3D applications, most publications are applied to 2D datasets. The method has been successfully applied in 2D and 3D using both synthetic and field data \citep{Ravasi2016target,neut2016adaptive, pereira2019internal,staring2019interbed,lomas2020marchenko,staring20203d}, however, the requirements of the reflection data are highly demanding, as the reflection data need to be well sampled in both time and space. This type of dense sampling is hard to obtain in the field for 3D acquisitions and furthermore, the 3D datasets are considerably larger in storage size than their 2D counterparts. \cite{lomas2020marchenko} made a comparison between results obtained with a full 3D acquisition and results obtained with linear seismic acquisition arrays, both recorded over a 3D medium. The authors showed that while the 2D approximation can yield good results, if one wants to take into account the full 3D effects, such as out-of-plane reflections, a 3D version of the Marchenko method is required.

The Marchenko equations can be solved in a variety of ways, the most common way is to use an iterative scheme \citep{Broggini2014}. However, the equations can also be solved using a least-squares inversion \citep{van2015inversion,Ravasi2017,Ravasi2021framework} or the iterative scheme can be combined with adaptive subtraction \citep{Costa2018laboratory,staring20203d}. The least-squares inversion is computationally feasible for 2D reflection data, however, for 3D data this method becomes computationally expensive. The adaptive subtraction is more robust to imperfections in the reflection data, however, due to its adaptive nature, the subtraction can attenuate physical events that are coinciding with multiples. 

Various open source implementations of the Marchenko method have been released over the years. The first of these implementations was written in the Madagascar environment, employing the 2D iterative scheme \citep{broggini2014data}. \cite{thorbecke2017implementation} published a C implementation of the Marchenko method that employs an iterative scheme for 2D reflection data that contain no free-surface multiples. More recently, an implementation in Python using the PyLops package was published that solves the method using least-squares inversion in 3D \citep{Ravasi2021framework}. In this paper, we extend the 2D iterative implementation of \cite{thorbecke2017implementation} to work with 3D reflection data. We discuss the practical challenges of the 3D extension, as the theoretical extension is straightforward, and show applications of the code.

We start by briefly considering the theory of the Marchenko method which forms the basis of the implementation. Next, we discuss the changes made to the 2D implementation to extend the implementation for 3D datasets. We make use of a floating point algorithm by \cite{Lindstrom2014}, called ZFP, to compress the reflection data to limit the required storage space and to reduce the loading time of the data. We also make use of a 3D Eikonal solver to obtain the first arrivals in the subsurface, to limit the modeling time that would be required if finite-difference modeling was used. We apply our 3D code to a subsection of the complex 3D Overthrust model by \cite{aminzadeh1997} for two applications. First, we retrieve Green's functions related to point sources inside the model and compare the results to a directly modeled reference. Next, we use these retrieved wavefields and focusing functions to apply structural imaging, based on the double-focusing approach \citep{staring2017adaptive} and show how the Marchenko method attenuates artifacts in the image, which are related to internal multiples. As this is a computationally expensive process in 3D, we demonstrate how the retrieval of the Green's functions can be performed for an entire depth level, through the use of plane-wave sources, based on the work by \cite{Meles2018}. This allows us to obtain a structural image of the subsurface much more efficiently, although only certain angles of reflectivity are obtained. We therefore retrieve the reflectivity for multiple angles in order to obtain a more complete structural image of the subsurface.

\section*{Theory}
Our 3D implementation of the Marchenko method follows the same approach as the 2D implementation by \cite{thorbecke2017implementation}. As the specifics of the implementation are described in great detail in that paper, we will only consider the most important equations, which form the basis of the implementation.

The Marchenko method allows one to retrieve Green's functions $G^{-,\pm}(\bxr,\bxa,t)$ in the subsurface, which are the impulse responses at point $\bxr$ at the surface of the Earth at time $t$ to a virtual point source at $\bxa$ in the subsurface. Note that we use a Cartesian coordinate system, where $\bx=(x_1,x_2,x_3)$ and $x_1$ and $x_2$ are the horizontal coordinates and $x_3$ is the depth coordinate. The first superscript in the Green's function indicates whether a wave is propagating downwards or upwards at $\bxr$, by using + or -, respectively, and the second superscript follows a similar notation for the radiation direction of the virtual source at $\bxa$. In order to retrieve the Green's functions, we require a reflection response $R(\bxr,\bxs,t)$, which is measured at location $\bxr$ at transparent surface $\mathbb{S}_0$, usually the surface of the Earth, as a result of a dipole source $\bxs$ at the same surface. The reflection response contains no free-surface multiples because of the assumption that $\mathbb{S}_0$ is transparent. Field datasets do contain this type of multiples, however, several robust techniques exist to remove these events \citep{verschuur1992adaptive,amundsen2001elimination,van2009estimating}. Alternatively, the Marchenko method can also be adjusted to directly work with data containing free-surface multiples \citep{singh2015marchenko}, however, this approach is not considered in this paper.

The Green's functions and reflection response can be related to each other via focusing functions $f_1^\pm(\bxr,\bxa,t)$, which are propagating downward or upward at $\bxr$, depending on the superscript. These focusing function focus from the surface to a focal point $\bxa$ in the subsurface, while correctly accounting for internal multiples. The relations between the Green's functions, focusing functions and reflection response are given as \cite[Equations (11) and (12)]{wapenaar2014marchenko},

\begin{align}
\label{Gf1a} 
G^{-,+}(\bxr,\bxa,t) + f_1^-(\bxr,\bxa,t) & =  \int_{\mathbb{S}_0} \int_{0}^\infty R(\bxr,\bxs,t') f_1^+(\bxs,\bxa,t-t') {\rm d}t' {\rm d} \bxs, \\
\label{Gf1b}
G^{-,-}(\bxr,\bxa,-t) + f_1^+(\bxr,\bxa,t) & =  \int_{\mathbb{S}_0} \int_{-\infty}^0 R(\bx,\bxs,-t') f_1^-(\bxs,\bxa,t-t') {\rm d}t' {\rm d} \bxs.
\end{align}

Note that these relations are based on the decomposed wavefields. The focusing functions can be combined together to create a total focusing function,
\begin{equation}
\label{f2decomp}
f_2(\bxa,\bxr,t) = f_1^+(\bxr,\bxa,t)-f_1^-(\bxr,\bxa,-t),
\end{equation}
where $f_2(\bxa,\bxr,t)$ is a focusing function that focuses from below to a focal point $\bxr$ at the surface. Similarly, following \cite{wapenaar2020reciprocity}, the decomposed Green's functions can be used to construct the total Green's function, $G(\bxr,\bxa,t)$,
\begin{equation}
\label{Gdecomp}
G(\bxr,\bxa,t) = G^{+,+}(\bxr,\bxa,t) + G^{+,-}(\bxr,\bxa,t) + G^{-,+}(\bxr,\bxa,t) + G^{-,-}(\bxr,\bxa,t).
\end{equation}
Note that if $\bxr$ is located at $\mathbb{S}_0$, the wavefield is purely upgoing, because the surface is transparent and the medium above the surface is homogeneous. In other words, $G^{+,+}(\bxr,\bxa,t) = G^{+,-}(\bxr,\bxa,t) = 0$ and the total Green's function only consists of $G^{-,+}(\bxr,\bxa,t)$ and $G^{-,-}(\bxr,\bxa,t)$, which are the Green's functions in \eqnsref{Gf1a} and \eqref{Gf1b}.

The Green's functions and focusing functions on the left hand side of \eqnsref{Gf1a} and \eqref{Gf1b} can be separated from each other based on causality relations. The direct arrival of the Green's function, $G_d(\bxr,\bxa,t)$, and the direct arrival of the downgoing focusing function, $f^+_{1,d}(\bxr,\bxa,t)$, are each other's inverse and therefore $f^+_{1,d}(\bxr,\bxa,-t)$ and $G_d(\bxr,\bxa,t)$ have the same arrival time $t_d(\bxr,\bxa)$, however, their amplitudes are different \citep{van2015green}. Furthermore, aside from the direct arrival, the entire wavefield of $G(\bxr,\bxa,t)$ arrives after $t_d(\bxr,\bxa)$ and the entire wavefield of $f_1^+(\bxa,\bxr,-t)$ and $f_1^-(\bxa,\bxr,t)$ arrives before $t_d(\bxr,\bxa)$, which means that the wavefields are separated in time, except for their direct arrival. Using these relations, we introduce an offset-dependent time-windowing function:
\begin{equation} \label{Thetadef}
\Theta(\bxr,\bxa,t)=\theta(t+(t_d(\bxr,\bxa)-\epsilon))-\theta(t-(t_d(\bxr,\bxa)-\epsilon))
\end{equation}
where $\theta(t)$ is the Heaviside step function and $\epsilon$ indicates a small constant, which is required to account for the band-limited nature of the wavefields. For simplicity, we will use the substitution $\Theta(\bxr,\bxa,t)=\Theta$. Applying $\Theta$ to the Green's function removes all events, however, when the window is applied to the focusing function only the first arrival is removed, while the coda will remain. Because the window is symmetric in time, it does not matter if the wavefields are time-reversed or not when the windowing function is applied. The window is applied to \eqnsref{Gf1a} and \eqref{Gf1b} in order to remove the Green's functions from the equations,

\begin{align}
\label{tf1a}
f_1^-(\bxr,\bxa,t)  & =  \Theta\int_{\mathbb{S}_0} \int_{0}^\infty R(\bxr,\bxs,t') f_1^+(\bxs,\bxa,t-t') {\rm d}t' {\rm d} \bxs, \\
\label{tf1b}
f_1^+(\bxr,\bxa,t) - f_{1,d}^+(\bxr,\bxa,t) & =  \Theta\int_{\mathbb{S}_0} \int_{-\infty}^0 R(\bxr,\bxs,-t') f_1^-(\bxs,\bxa,t-t') {\rm d}t' {\rm d} \bxs.
\end{align}

\eqnsref{tf1a} and \eqref{tf1b} are the coupled Marchenko equations. Note that in \eqnref{tf1b}, we subtract the direct arrival from the downgoing focusing function to account for the removal of this direct arrival by the windowing function. Assuming that the direct arrival is known, in this system we have two unknowns and two equations, allowing us to solve it in an iterative manner. The reflection response is the known quantity and will not change, while the upgoing and downgoing focusing function will be updated according to

\begin{align}
\label{tf1ak}
f_{1,k}^-(\bxr,\bxa,t) &  =  \Theta\int_{\mathbb{S}_0} \int_{0}^\infty R(\bxr,\bxs,t') f_{1,k}^+(\bxs,\bxa,t-t') {\rm d}t' {\rm d} \bxs, \\
\label{tf1bk}
f_{1,k+1}^+(\bxr,\bxa,t) & =  \Theta\int_{\mathbb{S}_0} \int_{-\infty}^0 R(\bxr,\bxs,-t') f_{1,k}^-(\bxs,\bxa,t-t') {\rm d}t' {\rm d} \bxs + f_{1,d}^+(\bxr,\bxa,t),
\end{align}

where $k$ indicates the iteration number. In \eqnref{tf1bk}, we added the direct arrival to both sides of the equation, so that the full downgoing focusing function is retrieved on the left hand side. To start the scheme, a first estimation is required. For the first estimation, we assume that the scattering coda of the downgoing focusing function is equal to zero, so that

\begin{align}
\label{ftm0}
f_{1,0}^+(\bxr,\bxa,t) = f^+_{1,d}(\bxr,\bxa,t).
\end{align}

As it is difficult to estimate $f^+_{1,d}(\bxr,\bxa,t)$, one can instead model $G_d(\bxr,\bxa,t)$, as it can be easily obtained using a background velocity model. $f^+_{1,d}(\bxr,\bxa,t)$ is often approximated using the final arrival of the time-reversed Green's function, $G_d(\bxa,\bxr,-t)$. As mentioned before, $f^+_{1,d}(\bxr,\bxa,t)$ and $G_d(\bxa,\bxr,-t)$ have the same arrival time, however, there is an amplitude difference. Using a time-reversed direct arrival instead of an inverted direct arrival as a first estimation of the focusing function will cause errors in the final result that are proportional to the transmission losses of the medium \citep{broggini2014data,brackenhoff2016rescaling,Neut2018single}. When $G_d(\bxr,\bxa,t)$ is modeled in a smooth version of the medium, instead of the exact medium, errors will be present in any case, which is why the time-reversal is more often used than the inversion of the direct arrival of the Green's function. Note that in imaging by deconvolution, discussed in the section Marchenko imaging, these amplitude errors should be compensated for.

\section*{Implementation}
\subsection*{Algorithm}
The implementation that is used in this paper makes use of a single compute kernel. By considering \eqnsref{tf1ak} and \eqref{tf1bk}, it can be seen that both of these equations make use of the same operation, namely a convolution of the reflection response or its time-reversal with either the downgoing or upgoing focusing function, followed by an application of the time window $\Theta$. It is more efficient to perform the convolution in the frequency domain, while the time window is applied in the time domain. We therefore consider these two operations separately in the kernel. Because the kernel is the same for both \eqnsref{tf1ak} and \eqref{tf1bk}, the equations can be written as a series expansion, similar to Equations (B-4) and (B-5) of \cite{wapenaar2014marchenko}

\begin{align}
\label{tf1ake}
-f_{1,k}^-(\bxr,\bxa,-t) & =  \sum_{i=0}^{k}N_{2i}(\bxa,\bxr,t), \\
\label{tf1bke}
f^+_{1,k+1}(\bxr,\bxa,t) & = \sum_{i=0}^{k}N_{2i+1}(\bxa,\bxr,t) + f_{1,d}^+(\bxr,\bxa,t) = \sum_{i=-1}^kN_{2i+1}(\bxa,\bxr,t),
\end{align}

where

\begin{align}
\label{Ni}
N_{i}(\bxa,\bxr,-t) & =  -\Theta RN_{i-1}(\bxa,\bxr,t),
\end{align}

\begin{align}
\label{RNi}
RN_{i-1}(\bxa,\bxr,\omega) & =  \sum^{n_S}_{j=1}R(\bxr,\bxjs,\omega) N_{i-1}(\bxa,\bxjs,\omega) \frac{\Delta x_{1}\Delta x_{2} \Delta t}{n_t},
\end{align}

\begin{align}
\label{N0}
N_{-1}(\bxa,\bxr,t) & =  f_{1,d}^+(\bxr,\bxa,t).
\end{align}

\eqnref{RNi} is performed in the frequency domain and therefore the functions depend on the angular frequency $\omega$ instead of time. The data are transformed to the frequency domain using the Fourier transform and as a result, the convolution between the reflection response and $N_i$ becomes a simple multiplication. To account for the discrete Fourier transform, the data needs to be divided by the number of time samples $n_t$ and a scaling with the temporal sampling $\Delta t$ is required for the numerical time convolution. We have also replaced the integral over the source locations by a numerical approximation, namely a sum over the source positions, where $\bxjs$ indicates the $j$-th source location of a total of $n_S$ source locations. Furthermore, for this approximation the data need to be scaled by the spatial sampling in the horizontal direction $x_1$ and $x_2$. $\Delta x_{1}$ and $\Delta x_{2}$ indicate the spatial sampling at $\bxjs$ in horizontal direction $x_1$ and $x_2$, respectively and can be different. For the 2D implementation, $\Delta x_{2}$ can be dropped and the scaling only depends on one horizontal direction.

\subsection*{3D challenges}
The basic algorithm that is described by \eqnsref{tf1ak}-\eqref{RNi} is the same as was used in the 2D implementation by \cite{thorbecke2017implementation} and the implementation is very similar, as the only addition is the integration along the second horizontal direction. However, there are additional complications that come with the 3D extension.

First, $R(\bxr,\bxs,\omega)$ has to be considered. The 3D reflection data must be pre-processed to comply with the assumptions made in the derivation of the Marchenko equations \cite[]{WapenaarJASA2014}. This processing has to at least include \cite[]{brackenhoff2019virtual};
\begin{itemize}
\item removal of free-surface multiples,
\item deconvolution with source wavelet,
\item crossline interpolation to avoid aliasing.
\end{itemize}
The removal of free-surface multiples and the deconvolution of the wavelet is beyond the scope of this paper, however, the crossline interpolation is a relevant topic as it relates to one of the most critical requirements of the reflection data, namely dense spatial sampling. Our implementation assumes that this requirement has been fulfilled, however, as a result the size of the reflection data will become very large. Loading the pre-processed reflection data from disk to memory is generally an inexpensive task in 2D, however, loading the full 3D reflection data matrix takes much longer and the storage space that is required for the data is much larger. To mitigate these problems a ZFP based compression algorithm is used \citep{Lindstrom2014}. The ZFP compression has been succesfully applied for the purpose of reducing the data size of seismic waveforms for waveform tomography with minimal errors \citep{lindstrom2016reducing}. Before the reflection data are compressed, they are transformed to the frequency domain, as the reflection data are only used in the frequency domain to circumvent the convolution, and only the data in the frequency-band of interest are compressed and stored to disk. Typically the lossless ZFP compression reduces the 3D data size by a factor of 4, which decreases the storage space of the data and the read-in time to memory. The program \path{TWtransform} (explained in appendix A.2) transforms uncompressed reflection time-data to the frequency domain, applies ZFP compression (based on tolerance) on a selected frequency range and writes the compressed data to disk. The data on disk contain a special compressed header that includes all location information present in the uncompressed Segy/SU headers that are needed in the 3D Marchenko program. The 3D Marchenko program has multiple options to read the reflection data, which can be done in the time-domain, frequency-domain, or compressed frequency-domain. In appendix A.1, the most important options and parameter settings of the 3D Marchenko program, \path{marchenko3D}, are explained in more detail.

Aside from the reflection data, the first estimation of the focusing function is also required, as dictated by \eqnref{N0}. To model this first arrival, a background velocity model is required. Using this model, the first arrival can be calculated using, for example, a finite-difference modeling code. In this paper, we make use of a 3D finite-difference modeling code \path{fdelmodc3D} that can be found in the OpenSource library by \cite{thorbecke2019opensource}. While the calculation of a single first arrival is feasible using a finite-difference code, it is very computationally expensive to calculate it for a large amount of focal points in this way. Furthermore, storing all these wavefields separately would require a large amount of storage space. As an alternative to the finite-difference modeling, an Eikonal, or ray-based solver can be employed. Using this solver, the first arrival of multiple focal locations can be efficiently computed and stored on disk. We employ the Eikonal solver \path{raytime3D} from the OpenSource library to calculate these first arrivals. This code is based on the theory by \cite{Vidale1990} and aside from the calculation of the arrival time, it also estimates a geometric spreading factor of the wavefield, based on the concepts by \cite{spetzler2005ray}. This amplitude estimation is important, as a uniform amplitude for all arrival times would cause significant artifacts in the result of the Marchenko method. This amplitude estimation is not exact and effects, such as transmission losses, are not taken into account, therefore there will be errors introduced in the amplitude of the estimated Green's function. The first arrival times that are estimated, either through the finite-difference modeling code or the Eikonal solver, are also used to construct the time window $\Theta$.

When the reflection data, time window and the first arrival are available, they can be used to compute the the focusing functions through the use of \eqnsref{tf1ak}-\eqref{RNi}. When the iterative scheme has converged to a solution, the final estimations of the upgoing and downgoing focusing functions can be used in \eqnsref{Gf1a} and \eqref{Gf1b} to compute the decomposed Green's functions. The \path{marchenko3D} program has the same functionality as its 2D counterpart \path{marchenko} to run multiple focal locations simultaneously, which means that the 3D reflection data only have to be read from the disk once. This is especially useful in 3D as loading the 3D reflection data from disk takes up a significant portion of time compared to the computing time of the Marchenko method.

\section*{Numerical Examples}
\begin{figure}[ht]
\includegraphics[trim={0cm 1cm 75cm 0cm},clip,width=\columnwidth]{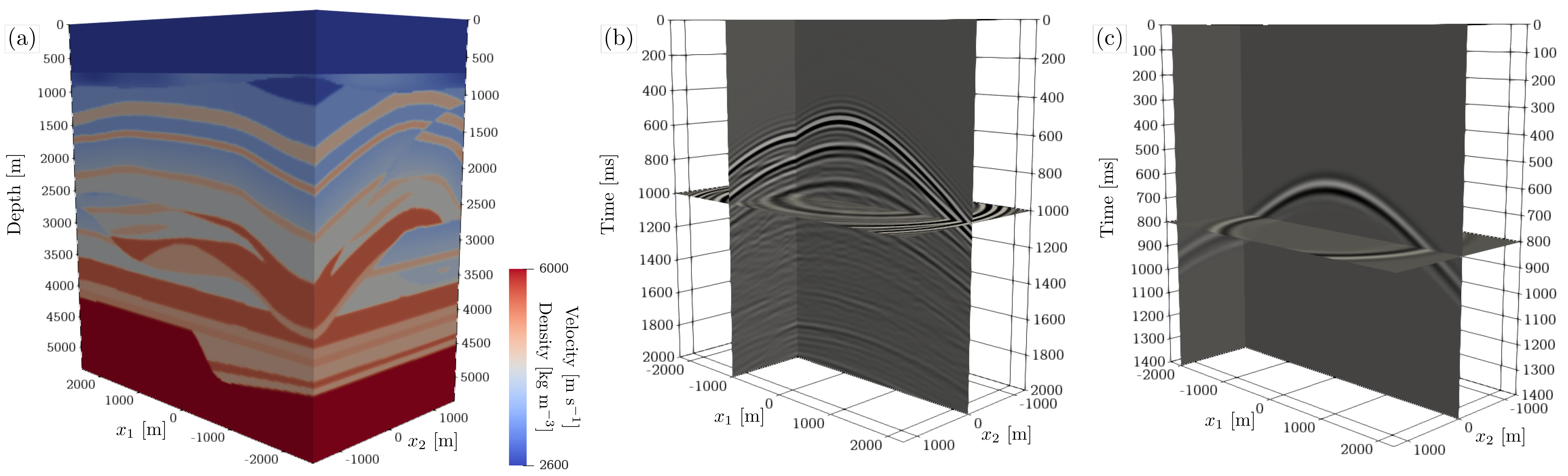}
\caption{(a) Subsection of the Overthrust model used for both the velocity and density contrasts. (b) A common-source record, with source position $\bxs=(0,0,0)$ and receivers at ${\bxr}$, containing a wavelet with a flat spectrum between 5 and 25Hz.}\label{fig:modelOV}
\end{figure}
We demonstrate the application of the 3D Marchenko scheme for two different applications, namely Green's function retrieval and imaging. We perform these applications for two types of virtual sources, namely point sources and plane-wave sources.

To take into account complex scattering in three dimensions, we demonstrate the method on numerical data that were modeled in a subsection of the SEG/EAGE Overthrust model from \cite{aminzadeh1997}, which is publicly available from the SEG Wiki (\url{https://wiki.seg.org/wiki/SEG/EAGE_Salt_and_Overthrust_Models}). We select a subsection of the model, as a recording setup over the full extent of the model is too large to fit in the memory of our computing nodes. Furthermore, this reduces the modeling time for the reflection response. We insert a layer with constant velocity and density above the model as a water layer, to simulate a marine setting. The velocity and density of the subsection are shown in Figure \ref{fig:modelOV}(a). The values of the density model are chosen to be the same as those of the velocity model to ensure strong reflections. We use the \path{fdelmodc3D} code to model the reflection response. An example of a shot record from a source at the surface in the center of the model is shown in Figure \ref{fig:modelOV}(b). For the full reflection response, we use a fixed spread acquisition, where the source is modeled at every receiver position. In the inline $x_1$-direction the sources and receivers are distributed from -2250 to 2250m with a spacing of 25m and in the crossline $x_2$-direction the acquisition ranges from -1250 to 1250m with a spacing of 50m. Note that the sampling distances in the inline and crossline directions are not equal, as is often the case for acquisition setups in the field. For the modeling of the reflection data, we apply a wavelet with a flat frequency spectrum that introduces ringing in the time-domain. This approximates the prerequisite of deconvolving for the source wavelet \citep{thorbecke2017implementation}. The frequency spectrum of this wavelet is flat between 5 and 25Hz, and tapered to zero with a cosine window from 5 to 0Hz and from 25 to 30Hz. The limited range of the frequency spectrum is chosen for modeling runtime purposes. The reflection data are modeled for 4.0s with a temporal sampling of 4ms. This temporal sampling complies with the industry standard, but note that a sampling of 16 ms would suffice. Before the reflection data are compressed, the size of the dataset is 387GB. After the frequency range is selected and the ZFP compression is applied, the size of the dataset is reduced to 46GB.

\subsection*{Green's function retrieval}
\begin{figure}[ht]
\centering
\includegraphics[width=1.0\columnwidth]{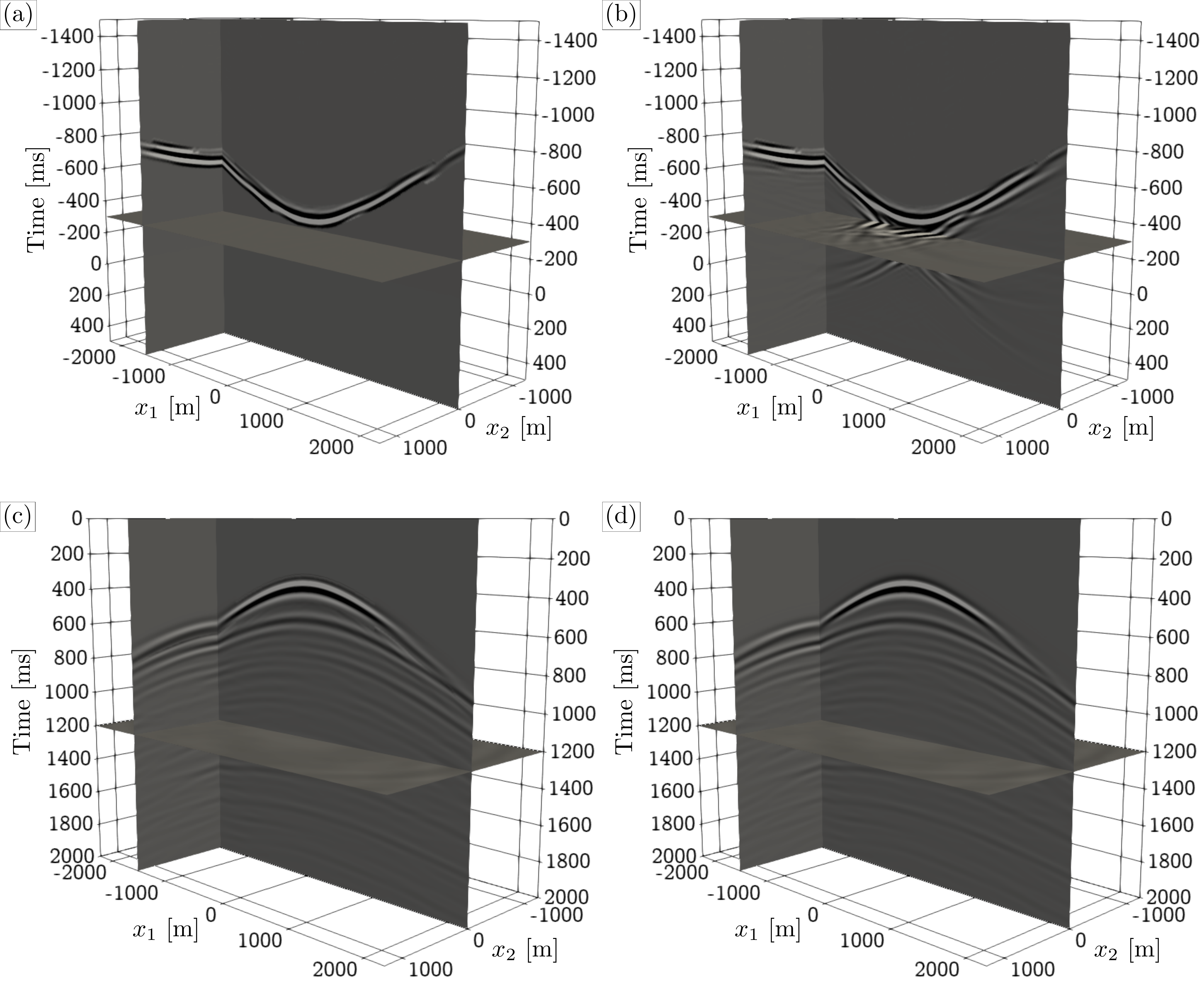}
\caption{(a) First arrival $f^+_{1,d}(\bxr,\bxa,t)$, modeled in the exact medium, (b) Focusing function $f_2(\bxa,\bxr,t)$ and (c) Green's function $G(\bxr,\bxa,t)$, both obtained through use of the Marchenko method and (d) Reference Green's function $G_{\rm ref}(\bxr,\bxa,t)$, modeled directly in the exact medium. All wavefields contain an 11Hz Ricker wavelet, are clipped at the same value and $\bxa=(0,0,1025)$m.}\label{functions}
\end{figure}

First, we consider the most basic application of the Marchenko method, namely, Green's function retrieval for a point source. To generate the necessary first arrival, we model a point source at $\bxa=(0,0,1025)$m using an 11Hz Ricker wavelet and record the response at the surface of the medium. We separate the first arrival from the coda of the wavefield to obtain $G_d(\bxr,\bxa,t)$. Instead of simply time-reversing this first arrival, we instead invert it to obtain the true $f^+_{1,d}(\bxr,\bxa,t)$, which is shown in \figref{functions}(a). In the field, it is unlikely that we would be able to achieve this, as one would need to have the exact model available, however, here we wish to demonstrate that the method can theoretically retrieve accurate amplitudes. We use the \path{marchenko3D} code to retrieve the decomposed focusing functions and Green's functions. We obtain the total focusing function $f_2(\bxa,\bxr,t)$ and the total Green's function $G(\bxr,\bxa,t)$ using \eqnsref{f2decomp} and \eqref{Gdecomp}, respectively and show these functions in \figref{functions}(b) and (c). 

\begin{figure}[ht]
\centering
\includegraphics[width=1.0\columnwidth]{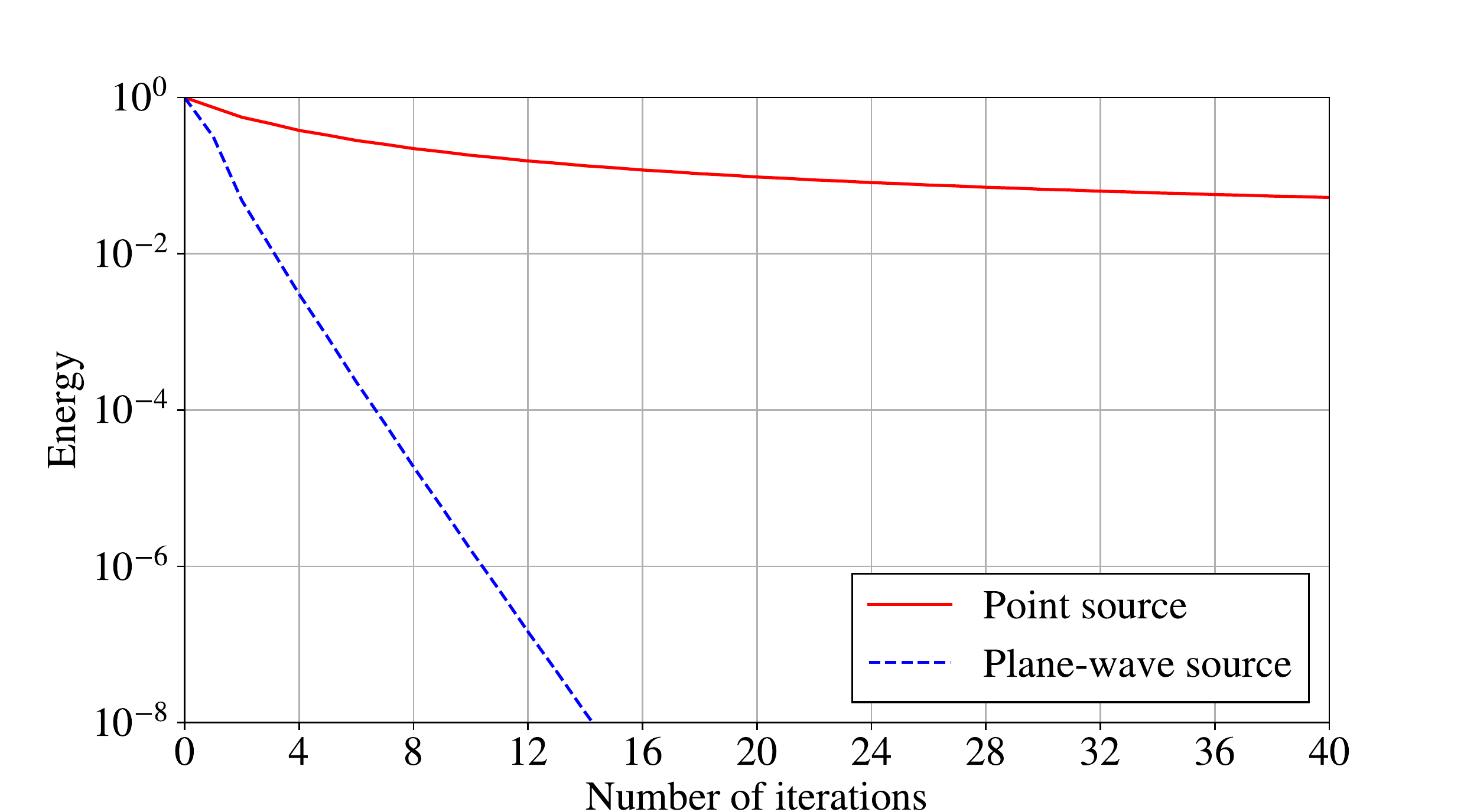}
\caption{Logarithmic convergence rate $\frac{|N_i|_{L_1}}{|N_0|_{L_1}}$ of the Green's function retrieval in the Overthrust model for a virtual point source at $\bxa=(0,0,1025)$m in solid red and a virtual plane-wave source for $\bpa=(80\mu$s m$^{-1},40\mu$s m$^{-1},1025$m$)$ in dashed blue. The y-axis is clipped at $10^{-8}$ as we do not expect anything below this energy level to significantly contribute to the final result.}\label{fig:ratecmplx}
\end{figure}

The convergence rate of the Marchenko method for a point source at $\bxa=(0,0,1025)$m is shown in \figref{fig:ratecmplx} as the solid red curve. This convergence rate is defined as the rate of the $L_1$ norm of the total energy in the update $N_i$ from \eqnref{Ni} to that of the $L_1$ norm of the total energy in the original update $N_0$. Note that a significant amount of iterations is required for convergence and that the convergence curve seems to flatten out instead of converging to zero. The low convergence rate has to do with the complexity of the model. Due to the fact that the model contains a large amount of reflectors and has a great variety of structures, the amount of iterations that are required is high. The maximum convergence is related to the aperture of the reflection response. The convergence to a relatively high energy level indicates that many limited-aperture artefacts are present in the update fields $N_i$. These limited-aperture effects are caused by the acquisition footprint, especially in the crossline direction. The complexity of the model causes scattering at large angles, which means that a larger aperture is required to properly capture all the scattered events. This is caused by the assumption of the Marchenko method that $\mathbb{S}_0$ extends infinitely in the horizontal directions, whereas in the field, it will always be limited.

To determine whether the retrieved Green's function is accurate, we compare it to a reference Green's function $G_{\rm ref}(\bxr,\bxa,t)$. This Green's function was obtained by modeling the source directly at $\bxa=(0,0,1025)$m and recording the response at the surface of the medium. It is shown in \figref{functions}(d). A visual inspection between the retrieved Green's function and the reference Green's function shows that the two wavefields appear to be similar, with a strong correlation between the arrival times and amplitudes of the events. The convergence to the correct solution is further supported by the decrease in energy per update, which can be interpreted from the solid red curve in \figref{fig:ratecmplx}.

\begin{figure}[ht]
\centering
\includegraphics[width=1.0\columnwidth]{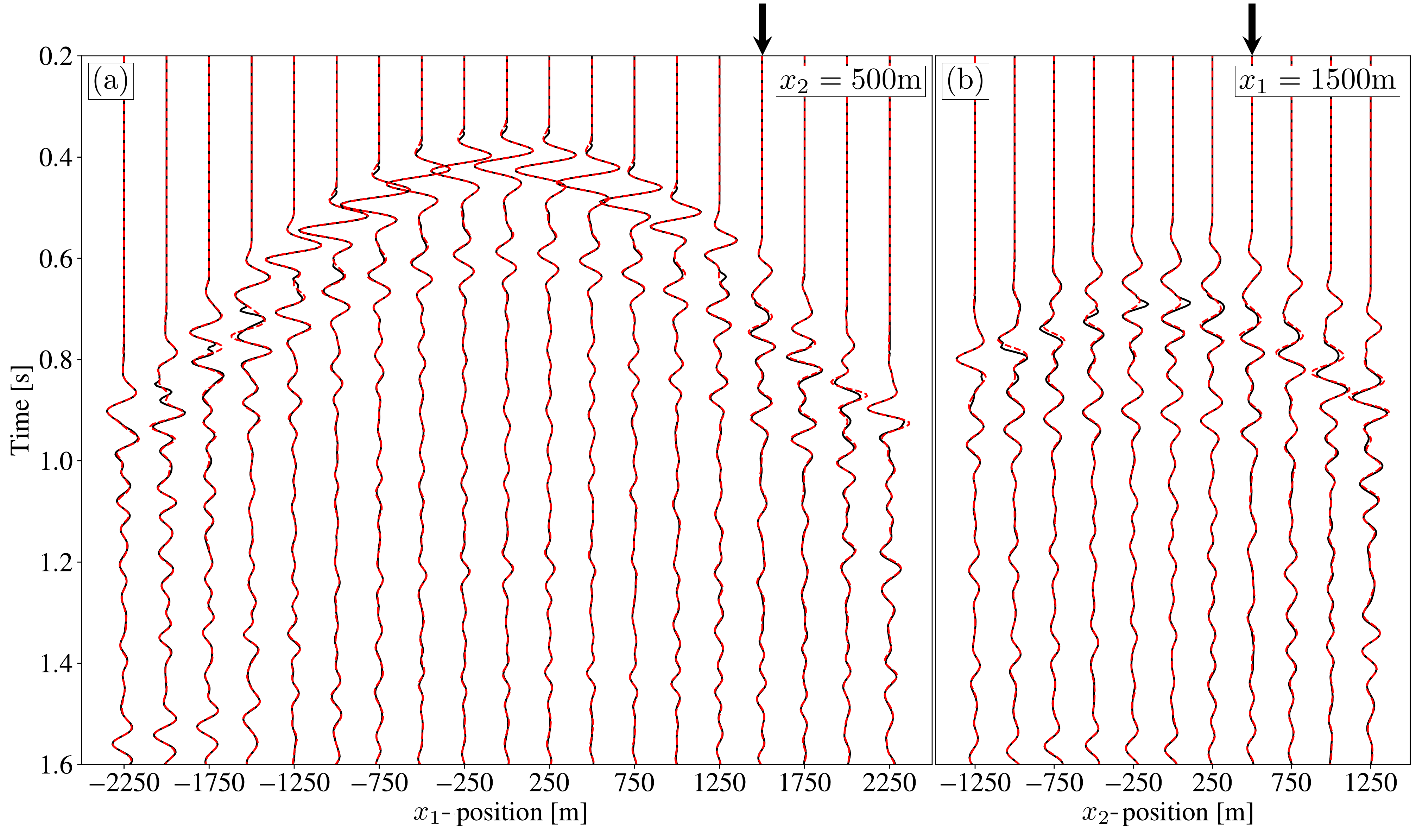}
\caption{Comparison between the reference Green's function $G_{\rm ref}(\bxr,\bxa,t)$ in dashed red and the Green's function $G(\bxr,\bxa,t)$ obtained through the use of the Marchenko method in solid black for (a) $x_2=$500m and (b) $x_1=$1500m and $\bxa=(0,0,1025)$m. All wavefields contain an 11Hz Ricker wavelet and have a gain applied of $t^{1.6}$ for display purposes. The black arrows on top indicate where the panels in (a) and (b) intersect.}\label{traces}
\end{figure}

To make a more accurate comparison, we show a direct comparison between some of the traces of the two functions in \figref{traces}. The traces of the retrieved Green's function are shown in black, while the traces of the reference Green's function are shown in dashed red. Note that in this display, we apply a time gain of $t^{1.6}$ to boost the amplitude of the events at later times. When the traces are compared, it can be seen that the match is excellent. The traces in \figref{traces}(a), which have a fixed offset in the crossline $x_2$ direction of 500m, are nearly identical for the two functions. There is a very slight error before the first event, which is caused by the window, and at high offsets from the source position the quality of the retrieved Green's function decreases slightly. This is because the events that have to be retrieved at this offset are not all captured by the recording aperture. This effect can be seen in \figref{traces}-(b), where the inline $x_1$ offset is fixed at 1500m. The errors are slightly larger here due to the fact that the entire panel is closer to the edge of the aperture. However, overall the match is still very strong and this demonstrates the capability of the Marchenko method to properly retrieve the Green's function from surface reflection data.

\subsection*{Marchenko imaging}
The Green's functions and focusing functions that are retrieved by using the Marchenko method can be employed for a variety of schemes. One of the most commonly used applications is imaging, where one aims to obtain the reflectivity of the subsurface in a region of interest. The reflectivity of the medium $\bar{R}_{\rm tar}(\bxb,\bxa,t)$ at a target depth is related to the decomposed Green's functions as \citep{amundsen2001elimination,wapenaar2014marchenko}
\begin{equation}
\label{Rtar}
G^{-,+}(\bxb,\bxr,t)=\int_{\mathbb{S}_A}\int_0^\infty \bar{R}_{\rm tar}(\bxb,\bxa,t-t')G^{+,+}(\bxa,\bxr,t')\text{d}t'\text{d}\bxb,
\end{equation}
where $\mathbb{S}_A$ is a surface inside the medium at the top of the target zone and the reciprocity relations $G^{-,+}(\bxb,\bxr,t)=G^{-,+}(\bxr,\bxb,t)$ and $G^{+,+}(\bxa,\bxr,t)=G^{-,-}(\bxr,\bxa,t)$ have been employed \citep{wapenaar2020reciprocity}. Through the use of a Multi-Dimensional Deconvolution (MDD) process, the reflectivity can be obtained from the decomposed Green's functions that are obtained using the Marchenko method. By employing this process, the overburden above the target depth is removed and only interactions from the target zone below $\mathbb{S}_A$ are preserved. Furthermore, if there are any errors present in the amplitude of $f_{1,d}^+$, the same errors will be present in both $G^{-,+}$ and $G^{-,-}$, so these errors will be removed for the larger part by the MDD process. However, properly applying the MDD, especially for 3D recording setups, is very difficult and prone to errors \citep{staring2017adaptive}. A more stable way of obtaining the reflectivity is through the use of the double-focusing method \citep{Neut2018single}. Here, no MDD is applied, and the decomposed focusing function is used. The double-focusing method can be expressed as
\begin{equation}
\label{doublefoc}
R_{\rm tar}(\bxb,\bxa,t)=\int_{\mathbb{S}_0}\int_0^\infty F^{+}(\bxr,\bxb,t')G^{-,+}(\bxr,\bxa,t-t')\text{d}t'\text{d}\bxr,
\end{equation}
with
\begin{equation}
\label{BFdef}
\dt F^+(\bxr,\bxb,t)=-\frac{2}{\rho} \partial_3f_1^{+}(\bxr,\bxb,t).
\end{equation}
In \eqnref{doublefoc}, $R_{\rm tar}(\bxb,\bxa,t)=G^{-,+}(\bxb,\bxa,t)$. Unlike the MDD approach, the double-focusing approach does not remove the overburden, but only redatums the sources and receivers from the surface of the medium to the target depth. Additionally, because there is no longer a deconvolution employed, the amplitude errors associated with $f_{1,d}^+$ are not removed. This means that we will not obtain the true amplitudes of the subsurface reflectors, however, we obtain the location of the reflectors, which means that we can retrieve an accurate structural image of the subsurface. The entire process is stable however, due to the fact that no deconvolution is employed. To obtain an image in a location inside the medium, without any influence of internal multiples, one can select the zero-time, zero-offset of the reflectivity in that location. This is defined as
\begin{equation}
\label{dfic}
r_{\rm im}(\bxb)=R_{\rm tar}(\bxb,\bxb,t=0).
\end{equation}
In this equation, the location of the virtual source and the virtual receiver coincides. For the imaging, only the first time sample is of interest. Hence, \eqnref{doublefoc} can be adjusted because the coda of the focusing function is meant to eliminate artifacts in the coda of the redatumed reflection response. Therefore, \eqnref{doublefoc} can be applied without the coda of the focusing function,
\begin{equation}
\label{doublefocd}
r_{\rm im}(\bxb)=R_{\rm tar}(\bxb,\bxb,t=0)=\int_{\mathbb{S}_0}\int_0^\infty F_d^{+}(\bxr,\bxb,t')G^{-,+}(\bxr,\bxb,-t')\text{d}t'\text{d}\bxr,
\end{equation}
with
\begin{equation}
\label{BFdefd}
\dt F_d^+(\bxr,\bxb,t)=-\frac{2}{\rho} \partial_3f_{1,d}^{+}(\bxr,\bxb,t).
\end{equation}
In \eqnref{doublefocd}, only the zero-time sample is retrieved, however, as stated before, the errors will be present in the coda of the redatumed reflection response and the zero time sample will be unaffected. The result can therefore also be used for the purpose of imaging without artifacts caused by internal multiples. Note that, similar to \eqnref{doublefoc}, any errors in $f_{1,d}^+$ are not accounted for. The double-focusing method has been successfully employed for the purpose of imaging using 3D acquisitions in the field, for an example, see \cite{staring20203d}.

To test the imaging potential of our \path{marchenko3d} program, we employ the double-focusing method. The required Green's functions and focusing functions are obtained by the Marchenko method, however, we make three alterations to the approach that was used for the Green's function retrieval. Firstly, instead of using the exact medium for estimating our first arrival, we apply a smoothing algorithm to create a background velocity model. Secondly, the first arrival is now modeled using a 3D Eikonal solver instead of using a finite-difference method. Finally, we no longer invert the first arrival, but simply take the time-reversal, as the changes to the modeling process of the first arrival will produce amplitude errors regardless. The changes in our approach are made to simulate field conditions and to keep the modeling time feasible. The Marchenko method needs to be performed for every location in the subsurface where an image is desired and therefore, the computational costs for producing an image are high.

Note that while the use of an Eikonal solver decreases the modeling time of the first arrival significantly, the use of finite-difference modeling would be a more accurate process, as an Eikonal solver does not represent the full physics of the wave equation. In media with strong velocity variations, the first arriving event may not be a purely propagating wavefield and may contain a refracted wave. The image obtained using an Eikonal solver will therefore be an approximation, however, it is computationally more feasible to apply the method like this in practice for a large amount of focal locations.

\begin{figure}[ht]
\centering
\includegraphics[width=\columnwidth]{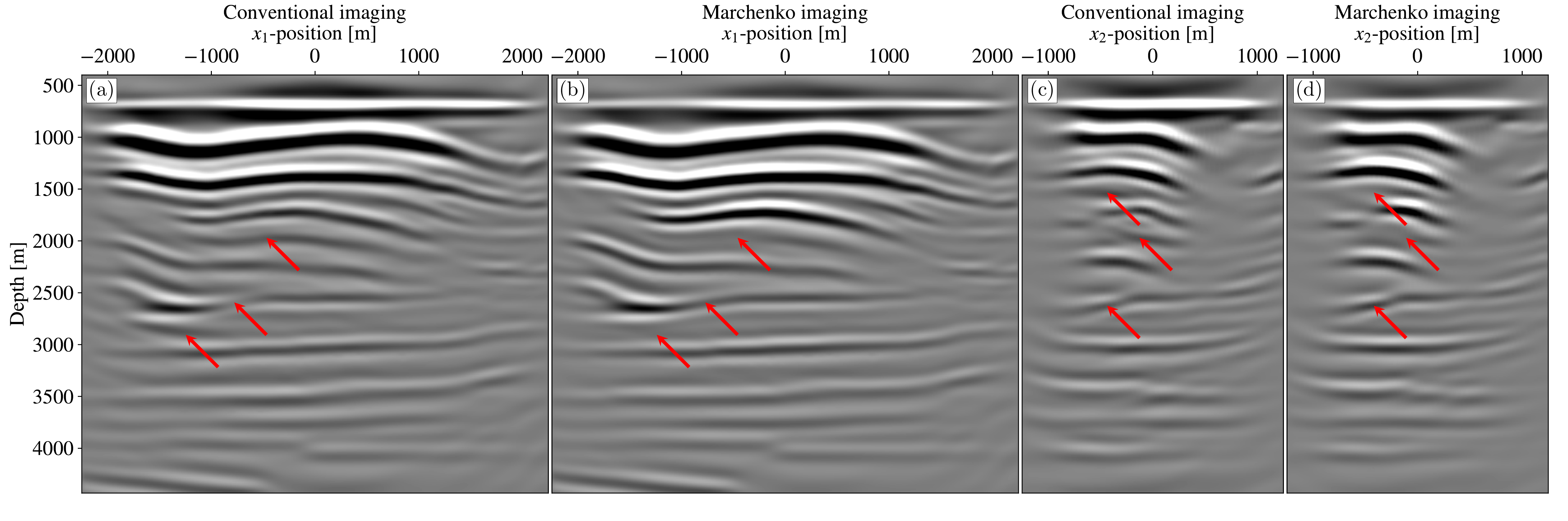}
\caption{Image of the Overthrust model along a fixed $x_2$ value of 0m using (a) conventional imaging and (b) Marchenko imaging after 30 iterations, and image of the Overthrust model along a fixed $x_1$ value of 0m using (c) conventional imaging and (d) Marchenko imaging after 30 iterations. The locations of artefacts that are attenuated by the Marchenko imaging are indicated by the red arrows.}\label{fig:cmplximag}
\end{figure}

We obtain the image of two cross sections of the Overthrust model, one inline cross-section for a fixed $x_2$-offset of 0m and one crossline cross-section for a fixed $x_1$-offset of 0m. The two cross-sections intersect each other in their respective centers. The focal locations are placed along a depth range of 400 to 4400m with a sampling of 25m. The range of the focal points in the inline direction is from -2250 to 2250m, with a sampling of 25m and in the crossline direction the range is set from -1250 to 1250m with a sampling distance of 50m. Similar to the Green's function retrieval, we perform 40 iterations for each focal point. After the focusing function and Green's function for each focal location are obtained using the Marchenko method and \eqnsref{Gf1a} and \eqref{Gf1b}, they are used in \eqnref{doublefoc} to obtain the local reflectivity. The zero time sample is then extracted to obtain the reflectivity at that exact location, following \eqnref{dfic}. For comparison, we also performed this approach using \eqnref{doublefocd} instead of \eqref{doublefoc}, however, for the imaging result, there were no differences, as is expected from the theory. Figure \ref{fig:cmplximag}(b) shows the reflectivity for each focal location for the inline direction and \ref{fig:cmplximag}(d) shows the same for the crossline direction. Conventional images for the inline and crossline directions are shown in Figures \ref{fig:cmplximag}(a) and \ref{fig:cmplximag}(c), respectively. As can be seen from the figures, the subsurface is complex and hard to resolve. Due to the small frequency bandwidth, the resolution of the images is limited, however, there are still artifacts present caused by the internal multiples, as indicated by the red arrows. The Marchenko imaging attenuates these artifacts, which shows that even on complex 3D models, our \path{marchenko3D} code can produce good results. The downside is that the computational costs remain high, even if compressed reflection data are used. Imaging all the points in the inline direction for a single depth level takes around 19800 seconds or five and a half hours, running in parallel on 40 Intel E5-2560 cores with a clock speed of 2.3GHz on a node using 256GB of 2133MHz RAM. Imaging the entire 3D medium using this approach may therefore prove too demanding unless a very powerful computing machine is available. A potential solution could be the use of GPUs for the implementation of the Marchenko method, which has shown promise for 2D Marchenko applications, for example in \cite{koehne2021multigpu}

\subsection*{Plane-wave Green's function retrieval}
Due to the high computational costs of creating 3D images in the subsurface, it is hard to efficiently apply imaging in practice using virtual point sources. As an alternative to the standard Marchenko imaging, one can image an entire depth level at once instead of just a single point. This can be achieved through the use of virtual plane-wave sources instead of virtual point sources. The idea of combining plane-waves with the Marchenko method was first proposed by \cite{Meles2018}, who applied the method with success in 2D settings. Here, we wish to implement the approach for 3D settings. The concept of the plane-wave method is that the focusing function no longer focuses to a single focal location in the subsurface, but rather to a focal depth. This gives a limited amount of information, due to the fact that only certain angles of the wavefield are considered. The plane-wave focusing functions $\tilde{f}^{\pm}_1(\bx,\bpa,t)$ are related to the focusing functions as
\begin{equation}
\label{fpw}
\tilde{f}^{\pm}_1(\bx,\bpa,t)=\int_{\mathbb{S}_A}f^{\pm}_1(\bx,\bxa,t-\bp\cdot\bx_{{\rm H},A}){\rm d}\bxa,
\end{equation}
where $\bx_{{\rm H},A}=(x_{1,A}-x_{1,c},x_{2,A}-x_{2,c})$, $\bp=(p_1,p_2)$ and $\bpa=(\bp,x_{3,A})$. Here, $p_1={\rm sin}\alpha{\rm cos}\beta/c$ and $p_2={\rm sin}\alpha{\rm sin}\beta/c$ are the horizontal ray parameters, where $\alpha$ is the dip angle, $\beta$ is the azimuth angle and $c=c(\bx)$ is the propagation velocity of the medium. $x_{1,c}$ and $x_{2,c}$ are the horizontal coordinates of the central location of the plane-wave source. By including these coordinates in the definition of $\bx_{{\rm H},A}$, we ensure that the center of the plane-wave source always has an emission time of $t=0$. \eqnref{fpw} states that the plane-wave focusing function $\tilde{f}^{\pm}_1$ is the integral of the focusing functions ${f}_1^{\pm}$ over all possible focal points $\bxa$ at $\mathbb{S}_A$ for a certain dip and azimuth given by the ray parameters. As such, $\tilde{f}^{\pm}_1(\bx,\bpa,t)$ does not focus to a single location, but instead focuses as a dipping wave, dictated by $\bp$, to a single depth level $x_{3,A}$. In case the ray parameters are zero, this wave is horizontal instead of dipping.

We also define the plane-wave Green's functions $\tilde{G}^{-,\pm}(\bxr,\bpa,t)$ which are related to the Green's functions
\begin{equation}
\label{Gpw}
\tilde{G}^{-,\pm}(\bxr,\bpa,t)=\int_{\mathbb{S}_A}G^{-,\pm}(\bxr,\bxa,t-\bp\cdot\bx_{{\rm H},A}){\rm d}\bxa.
\end{equation}
The plane-wave Green's functions behave similarly to the regular Green's functions, however, instead of being the impulse response of a point source, the plane-wave Green's functions are the response of a medium to a plane-wave source at $x_{3,A}$, which generates a dipping wave as dictated by the ray parameters.

The plane-wave versions of the focusing functions and the Green's functions can be combined with the \eqnsref{Gf1a} and \eqref{Gf1b} simply by applying the integration over all possible focal points for a set of ray parameters

\begin{align}
\label{Gf1apw} 
\tilde{G}^{-,+}(\bxr,\bpad,t) + \tilde{f}_1^-(\bxr,\bpad,t) & =  \int_{\mathbb{S}_0} \int_{0}^\infty R(\bxr,\bxs,t') \tilde{f}_1^+(\bxs,\bpad,t-t') {\rm d}t' {\rm d} \bxs, \\
\label{Gf1bpw}
\tilde{G}^{-,-}(\bxr,\bpa,-t) + \tilde{f}_1^+(\bxr,\bpad,t) & =  \int_{\mathbb{S}_0} \int_{-\infty}^0 R(\bx,\bxs,-t') \tilde{f}_1^-(\bxs,\bpad,t-t') {\rm d}t' {\rm d} \bxs.
\end{align}

In these equations, we define $\bpad=(-\bp,x_{3,A})$, which all plane-wave wavefields contain with the exception of $\tilde{G}^{-,-}(\bxr,\bpa,-t)$. This is due to the fact that this particular wavefield is time-reversed in \eqnref{Gf1b}. The change to plane-waves has not affected the reflection response, only the focusing functions and Green's functions, which means that no changes to the reflection response have to be made. Similarly to the original equations, the Green's functions in the system of \eqnsref{Gf1apw} and \eqref{Gf1bpw} can be suppressed through the use of a windowing function. The change to plane-waves does affect the causality relations however, especially because $\tilde{G}^{-,-}(\bxr,\bpa,t)$ and $\tilde{G}^{-,+}(\bxr,\bpad,t)$ are dipping at different angles. Hence, the time window needs to be adjusted so the dipping plane-waves are properly handled \citep{meles2020data}. These windows are defined as
\begin{equation}\label{Thetapwdef}
\begin{split}
\tilde{\Theta}=\tilde{\Theta}(\bxr,\bpa,t)&=\theta(t-t_a)-\theta(t-t_b),\\
t_a&=-\tilde{t}_d(\bxr,\bpa)+\epsilon,\\
t_b&=\tilde{t}_d(\bxr,\bpad)-\epsilon,
\end{split}
\end{equation}
where $\tilde{t}_d(\bxr,\bpad)$ is the first possible arriving event of $\tilde{G}^{-,+}(\bxr,\bpad,t)$ and $-\tilde{t}_d(\bxr,\bpa)$ is the last arriving event of $\tilde{G}^{-,-}(\bxr,\bpa,-t)$. As a result, the windowing function is no longer symmetrical in time, instead, the window has parallel planes. For a further derivation of this window, see \cite{wapenaar2021overview}. Applying this window to \eqnsref{Gf1apw} and \eqref{Gf1bpw} results in the plane-wave coupled Marchenko equations

\begin{align}
\label{tf1apw}
\tilde{f}_1^-(\bxr,\bpad,t)  & =  \tilde{\Theta}\int_{\mathbb{S}_0} \int_{0}^\infty R(\bxr,\bxs,t') \tilde{f}_1^+(\bxs,\bpad,t-t') {\rm d}t' {\rm d} \bxs, \\
\label{tf1bpw}
\tilde{f}_1^+(\bxr,\bpad,t) - \tilde{f}_{1,d}^+(\bxr,\bpad,t) & =  \tilde{\Theta}\int_{\mathbb{S}_0} \int_{-\infty}^0 R(\bxr,\bxs,-t') \tilde{f}_1^-(\bxs,\bpad,t-t') {\rm d}t' {\rm d} \bxs.
\end{align}

Note that \eqnsref{tf1apw} and \eqref{tf1bpw} have a very similar structure to \eqnsref{tf1a} and \eqref{tf1b}. They can therefore be solved with the same Marchenko method as was used for the point source focusing functions. The only things that have to be adjusted are the input and the windowing function. The first estimation of the focusing function can once again be estimated by using the direct arrival

\begin{align}
\label{ftm0pw}
\tilde{f}_{1,0}^+(\bxr,\bpad,t) = \tilde{f}^+_{1,d}(\bxr,\bpad,t).
\end{align}

However, because we cannot compute this first arrival directly, we have to model the direct arrival of the Green's function and time-reverse it. The plane-wave is modeled by placing point sources along a horizontal plane in the medium of interest and emitting with a time delay, so that the emission time of a location along the horizontal plane is $\delta(t-\bp\cdot\bx_{{\rm H},A})$. By taking the direct arrival of the Green's function, we obtain $\tilde{G}^+_{d}(\bxr,\bpa,t)$, which we can then relate to the direct arrival of the focusing function as

\begin{align}
\label{fgrelpw}
\tilde{f}^+_{1,d}(\bxr,\bpad,t) &= \tilde{G}_{d}(\bxr,\bpa,-t).
\end{align}

Notice that according to \eqnref{fgrelpw}, the original ray parameters that were used to model the plane-wave source and obtain $\tilde{G}_{d}(\bxr,\bpa,t)$ are the opposite of the ray parameters that are included in the first estimation of the focusing function $\tilde{f}^+_{1,d}(\bxr,\bpad,t)$. Hence, when we use \eqnref{Gf1bpw} to obtain $\tilde{G}^{-,-}(\bxr,\bpa,t)$, it is dipping according to the original ray parameters and when we use \eqnref{Gf1apw} to obtain $\tilde{G}^{-,+}(\bxr,\bpad,t)$ it is dipping according to the opposite of the original ray parameters.

\begin{figure}[ht]
\centering
\includegraphics[width=1.0\columnwidth]{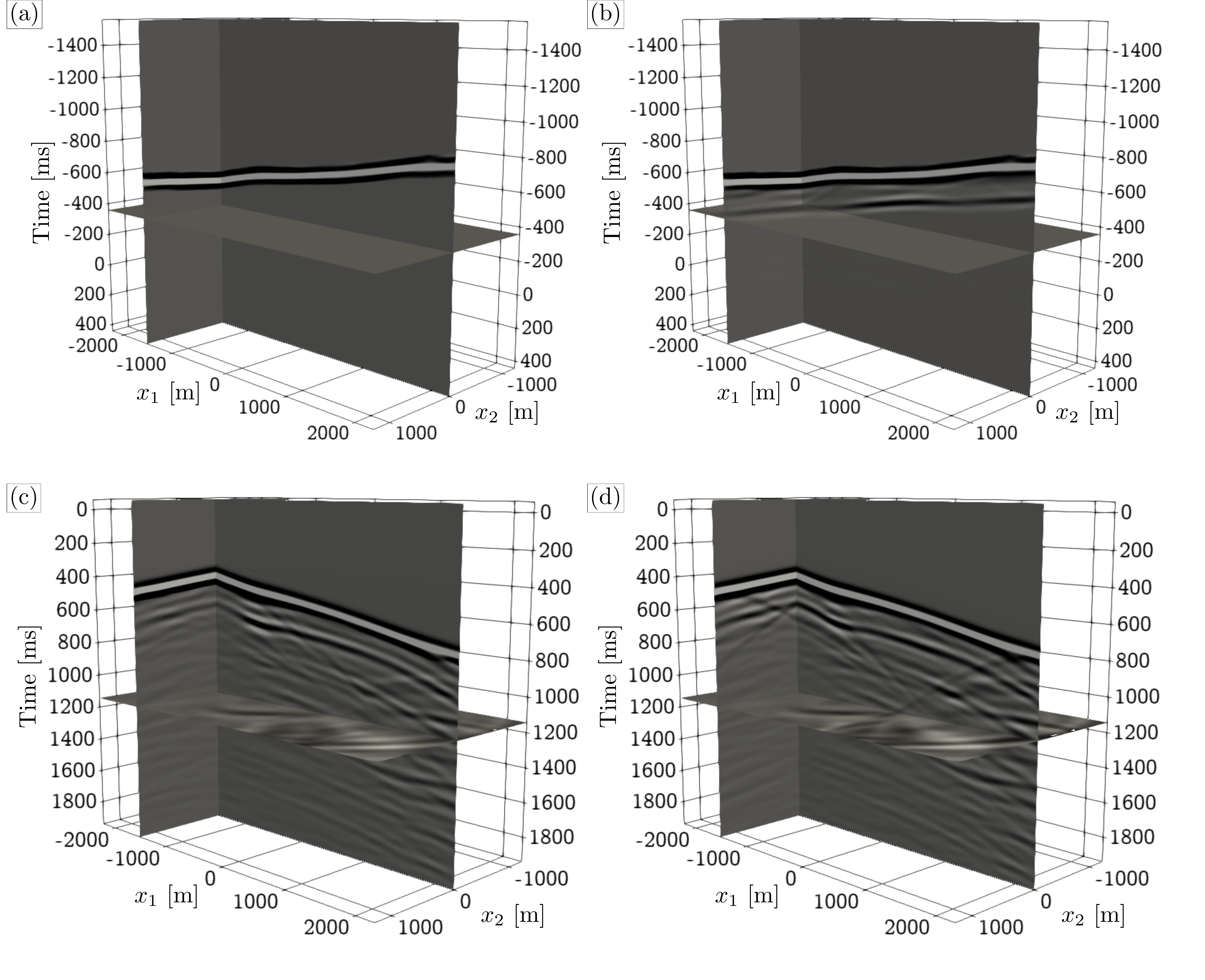}
\caption{(a) First arrival $\tilde{f}^+_{1,d}(\bxr,\bpad,t)$, modeled in the exact medium, (b) Focusing function $\tilde{f}_2(\bpad,\bxr,t)$ and (c) Green's function $\tilde{G}(\bxr,\bpa,t)$, both obtained through use of the Marchenko method and (d) Reference Green's function $\tilde{G}_{\rm ref}(\bxr,\bpa,t)$, modeled directly in the exact medium. All wavefields contain an 11Hz Ricker wavelet, are clipped at the same value and $\bpa=(80\mu$s m$^{-1},40\mu$s m$^{-1},1025$m$)$.}\label{functions_plane}
\end{figure}

We retrieve the Green's function in the subsurface at a depth level of $x_3=1025$m. We use a plane-wave that is dipping in both the inline and the crossline direction as determined by ray parameters $\bp=(80,40)\mu$s m$^{-1}$. Because the velocity of the medium varies at this depth, we cannot convert this to a single angle, however, the majority of this depth level corresponds to a velocity of 4500 m s$^{-1}$. Using this velocity we calculate a representative dip and azimuth angle, namely $\alpha=23.6^{\circ}$ and $\beta=26.6^{\circ}$. In order to obtain the full Green's function, we run the Marchenko method twice, once using $\bp$ to obtain $\tilde{G}^{-,-}(\bxr,\bpa,t)$ and once using $-\bp$ to obtain $\tilde{G}^{-,+}(\bxr,\bpa,t)$ , and use these results in \eqnref{Gdecomp}. The initial estimation of the focusing function $\tilde{f}^+_{1,d}(\bxr,\bpad,t)$, which we obtained in the exact model for accurate amplitudes, is shown in \figref{functions_plane}(a), while the full focusing function $\tilde{f}_{2}(\bpad,\bxr,t)$ is shown in (b). To construct this focusing function, we apply the plane-wave definition to \eqnref{f2decomp},
\begin{equation}
\label{f2decomppw}
\tilde{f}_2(\bpad,\bxr,t) = \tilde{f}_1^+(\bxr,\bpad,t)-\tilde{f}_1^-(\bxr,\bpa,-t).
\end{equation}
The convergence of the Marchenko method using $\bp$ is shown as the dashed blue curve in \figref{fig:ratecmplx}. If $-\bp$ is used, the graph is visually identical. Note that the method converges much faster if a plane-wave source is used instead of a point source, this is caused by the fact that only certain angles are covered by the dipping plane-wave, hence the amount of propagation directions that needs to be resolved is much lower. The full Green's function $\tilde{G}(\bxr,\bpa,t)$, which is obtained through the use of the two runs of the Marchenko method is shown in \figref{functions_plane}(c).

We compare the retrieved Green's function to a reference Green's function $\tilde{G}_{\rm ref}(\bxr,\bpa,t)$, that was directly modeled in the medium and is shown in \figref{functions_plane}(d). Visually, the two Green's functions appear to be very similar. There are some slight differences caused by curving events originating from the edges of the aperture. These are caused by the modeling of the plane-wave source as opposed to a point source. The amplitude along the wavefront is nearly constant for a plane-wave source, while for a point source it decreases near the edges. Therefore, the edge effects are more pronounced for the plane-wave sources. When the first arrival is isolated from the coda, these effects are largely removed, because they are present in the coda, hence the difference in the results.

\begin{figure}[ht]
\centering
\includegraphics[width=1.0\columnwidth]{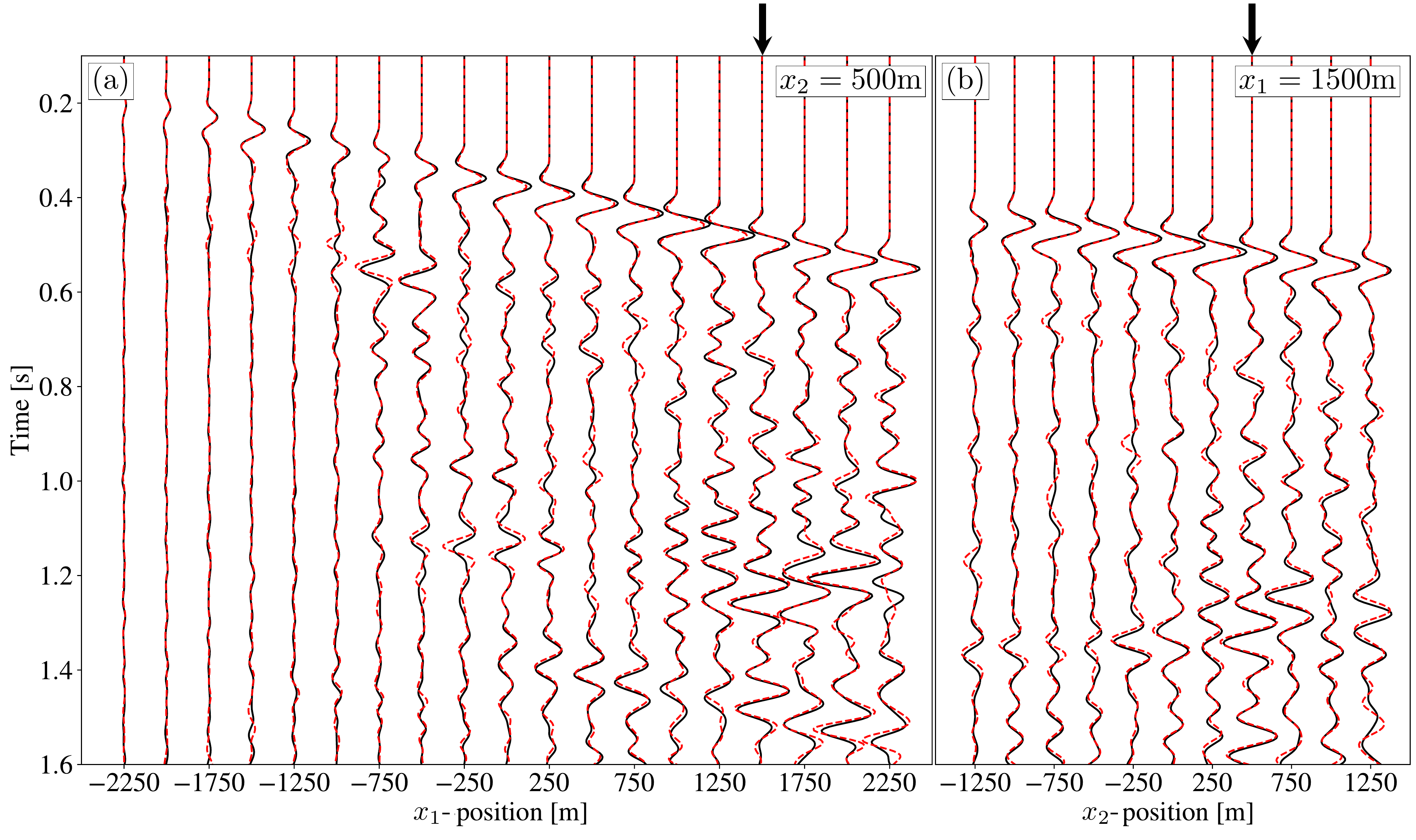}
\caption{Comparison between the reference Green's function $\tilde{G}_{\rm ref}(\bxr,\bpa,t)$ in dashed red and the Green's function $\tilde{G}(\bxr,\bpa,t)$ obtained through the use of the Marchenko method in solid black for (a) $x_2=$500m and (b) $x_1=$-1500m and $\bpa=(80\mu$s m$^{-1},40\mu$s m$^{-1},1025$m$)$. All wavefields contain an 11Hz Ricker wavelet and have a gain applied of $t^{1.6}$ for display purposes. The black arrows on top indicate where the panels in (a) and (b) intersect.}\label{traces_plane}
\end{figure}

To further investigate the accuracy of the retrieved Green's function, we construct a similar figure as \figref{traces}, where we display traces of the two Green's function next to each other. The result is shown in \figref{traces_plane}. The accuracy has decreased in comparison with the standard point-source Green's function retrieval, which is also a result from the plane-wave setup. The errors caused by placing the source near the edges contribute more to the final result than when point sources are used, hence the decrease in quality. Overall, the match is still very strong, for both the amplitude and the phase of the events.

\subsection*{Plane-wave imaging}
We perform plane-wave imaging through the double-focusing method, similar to \eqnref{doublefoc}. We apply the integral over all possible focal points, following \eqnsref{fpw} and \eqref{Gpw},
\begin{equation}
\label{dfpw}
\tilde{R}_{\rm tar}(\bxb,\bpa,t)=\int_{\mathbb{S}_0}\int_0^\infty F^{+}(\bxr,\bxb,t')\tilde{G}^{-,+}(\bxr,\bpa,t-t')\text{d}t'\text{d}\bxr.
\end{equation}
In \eqnref{dfpw}, $\tilde{G}^{-,+}$ and $\tilde{R}_{\rm tar}$ are both plane-wave responses, while $F^{+}$ still is the focusing function related to a focal point instead of a focal depth. We can use the redatumed receiver response to formulate an image condition for each location in the surface
\begin{equation}
\label{dficpw}
\tilde{r}_{\rm im}(\bxb,\bpa)=\tilde{R}_{\rm tar}(\bxb,\bpa,t=\bp\cdot\bx_{{\rm H},B}).
\end{equation}
Here, $\tilde{r}_{\rm im}(\bxb,\bpa)$ is the reflectivity in the subsurface at location $\bxb$ given a certain angle dictated by $\bpa$. The time sample that is selected from the redatumed receiver response is changed, depending on the angle of the incoming wave, which is required due to the time-delay associated with the angle of the wavefront. By choosing different values for $\bpa$, a reflection image can be obtained for various angles.

Note that in order to obtain the correct angle-dependent reflectivity, both $F^{+}(\bxr,\bxb,t)$ and $\tilde{G}^{-,+}(\bxr,\bpa,t)$ need to have accurate amplitudes. As we mentioned, this is hard to achieve using the Marchenko method in practice. As a result, we will not retrieve the true angle-dependent reflectivity of the subsurface, however, we can obtain a structural image of the subsurface related to specific ray parameters. In order to obtain the true angle-dependent reflectivity, \eqnref{dfpw} can be rewritten as a deconvolutional process, similar to \eqnref{Rtar}, but this is beyond the scope of this paper, as we are interested in efficiently obtaining a structural image of the subsurface.

The procedure using \eqnref{dfpw} has a notable disadvantage in that not only the plane-wave response needs to be obtained, but also the focusing function for each focal point in the subsurface, which means that the computational effort compared to using \eqnref{doublefoc} in fact increases. However, if we instead use an approach similar to \eqnref{doublefocd}, we can avoid the computation of the full focusing function. We rewrite \eqnref{dfpw} combined with \eqref{dficpw} as
\begin{equation}
\label{dfpwd}
\tilde{r}_{\rm im}(\bxb,\bpa)=\tilde{R}_{\rm tar}(\bxb,\bpa,t=\bp\cdot\bx_{{\rm H},B})=\int_{\mathbb{S}_0}\int_0^\infty F_d^{+}(\bxr,\bxb,t')\tilde{G}^{-,+}(\bxr,\bpa,\bp\cdot\bx_{{\rm H},B}-t')\text{d}t'\text{d}\bxr.
\end{equation}
We only model the first arrival for each point we want to image and do not need to apply the Marchenko method to obtain the full focusing function, saving considerable computational costs. 

\begin{figure}[ht]
\centering
\includegraphics[width=\columnwidth]{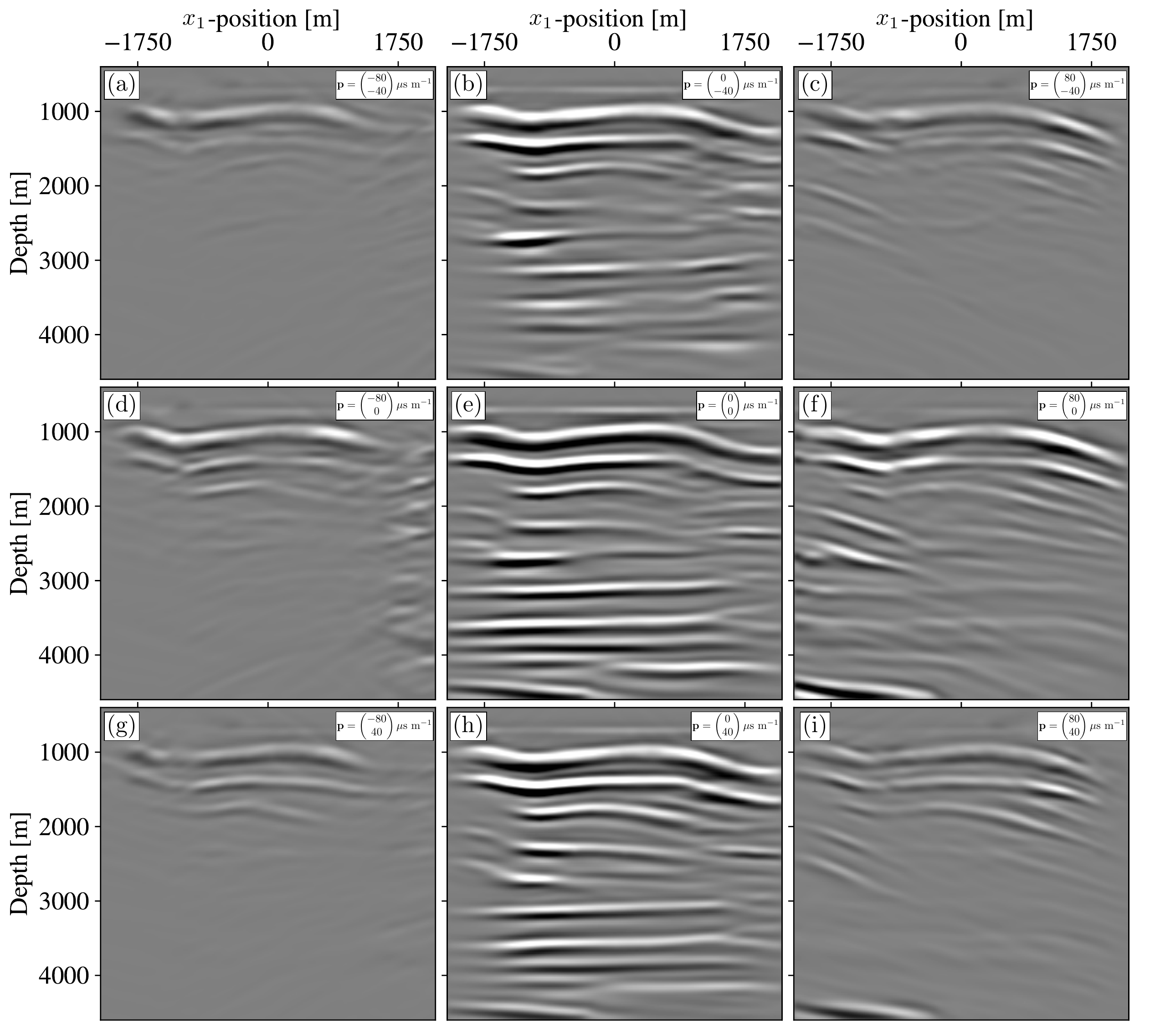}
\caption{Angle-dependent reflectivity of the subsurface obtained through the use of Marchenko imaging for different ray parameters.}\label{imagangles}
\end{figure}

To demonstrate the results of the method, we apply \eqnref{dficpw} to the Overthrust model. We retrieve a horizontal plane-wave, i.e. $\bp=(0,0)$, and image the same section as we did in \figref{fig:cmplximag}(b). The first arrival for the plane-wave $\tilde{f}^+_{1,d}$ is obtained through the use of finite-difference modeling, while the first arrival $f^+_{1,d}$ that is used in \eqnref{BFdefd} and subsequently in \eqnref{dfpwd} is obtained using the Eikonal solver. The result is shown in \figref{imagangles}(e). For comparison, for the standard Marchenko image we perform the Marchenko method for 161 depths and 181 lateral positions, for a total of 29141 different points, while for the plane-wave imaging, we perform the Marchenko method only for the 161 depths. The computation times for all these depths using plane-wave sources is around five and a half hours, which is similar to the computational time of all focal points in the inline direction for a single depth when the standard Marchenko imaging is employed. Because the first arrivals that are used for the imaging step are all obtained using an Eikonal solver, the computational costs of retrieving these events are negligible compared to the computational costs of the Marchenko method. The image that is obtained using the horizontal plane-wave contains many similar reflectors as were seen in the standard Marchenko image in \figref{fig:cmplximag}(b). However, because we are only considering reflectivity that is related to certain ray parameters, the reflectors are not as well resolved as the standard Marchenko image. 

We consider additional angles of reflectivity in order to resolve the events in the plane-wave image more clearly \citep{Meles2018}. These images are shown in \figref{imagangles}, for varying ray parameters. The ray parameter in the inline direction takes values of -80, 0 and 80$\mu$s m$^{-1}$ and the ray parameter in the crossline direction takes values of -40, 0 and 40$\mu$s m$^{-1}$. The central column shows the result when the inline ray parameter is zero and we can see that these panels contain the most energy. This indicates that a large amount of the reflectors have a small dipping angle in the inline direction. The deepest part of the image also contains less energy when the crossline ray parameter is negative rather than positive, which indicates that the reflectors at this depth are dipping positively in the crossline direction. The images constructed with a negative ray parameter in the inline direction contain considerably less energy than when the ray parameter is positive. The energy is mostly constrained to the shallow part of the image and the edges. Especially near the edges it is hard to determine whether this is caused by the edge of the aperture or by genuine reflector direction. The images that have a dip in the inline direction contain more energy and have reflectors at various depths and at distances away from the edge. 

\begin{figure}[ht]
\centering
\includegraphics[width=\columnwidth]{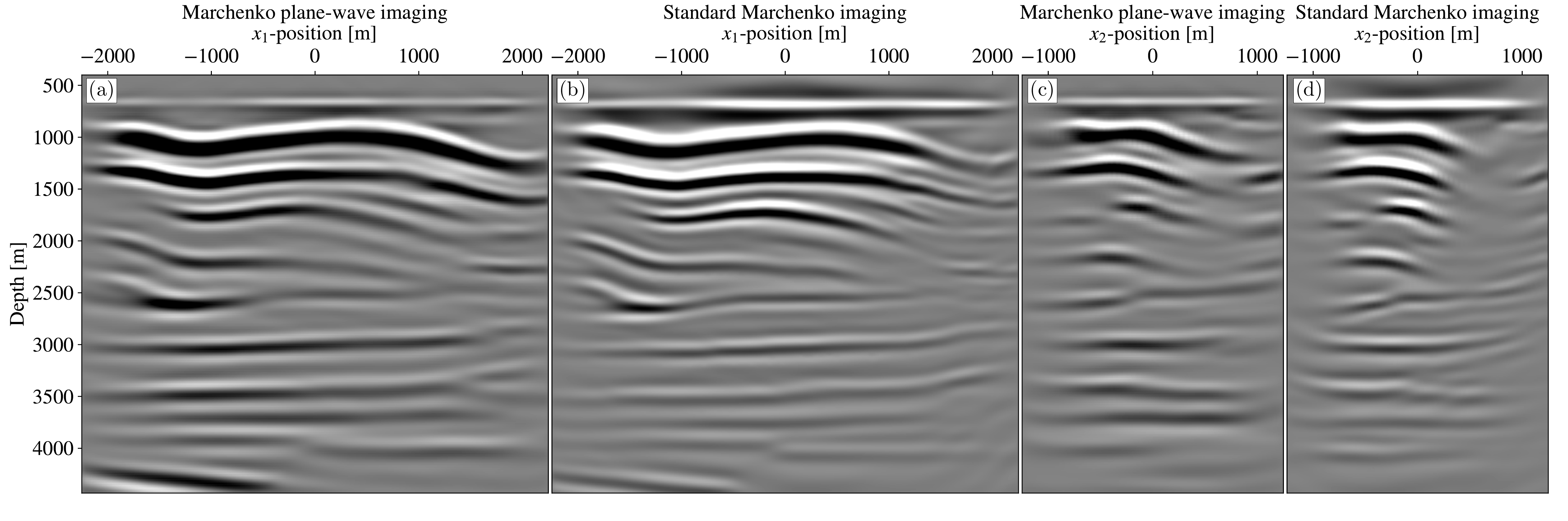}
\caption{Image of the Overthrust model along a fixed $x_2$ value of 0m using (a) plane-wave and (b) standard Marchenko imaging after 30 iterations, and image of the Overthrust model along a fixed $x_1$ value of 0m using (c) plane-wave and (d) standard Marchenko imaging after 30 iterations. (a) is constructed by combining the plane-wave images in \figref{imagangles} and (c) is constructed in a similar way.}\label{fig:planeimag}
\end{figure}

\figref{imagangles} gives insight into the reflector dip of the subsurface. We combine these results together to create a single image, which is shown in \figref{fig:planeimag}(a). For comparison, we plot the result of the standard Marchenko image in (b).  Most of the major reflectors are present at the same locations and with similar reflector strength in both imaging results. The result of the standard Marchenko imaging is better resolved, because this approach takes into account all angles of reflections within the range that is measured by the aperture. While the result of the plane-wave images is of a lesser quality, the computational costs are significantly lower, even if several images associated with different ray parameters have to be retrieved. We repeat the retrieval for both the standard Marchenko approach and plane-wave Marchenko approach to retrieve the image in the crossline direction. The result for the plane-wave Marchenko image is shown in \figref{fig:planeimag}(c) and for the standard Marchenko image in (d). From these images we can draw similar conclusions as we did from the images in the inline direction. The same events are resolved however, the resolution of the standard Marchenko image is higher than that of the plane-wave image. It should also be noted that the plane-wave image does not contain any of the artifacts that are present when conventional imaging is used, as the plane-wave Marchenko method still attenuates all the internal multiple artifacts.

\section*{Discussion}
Our examples have all been performed using synthetic data. For field data, additional complications would arrive, for example, subsampling in space of the reflection data. The crossline spacing of the sources and receivers tends to be much sparser than the inline spacing, so some kind of interpolation needs to be applied \citep{staring2020narrow}. In case the data have irregular sampling, the standard Marchenko scheme can be adjusted to handle this \citep{ijsseldijk2020adaptation,haindl2021handling}. In case the data are well sampled, it has been shown that the size of the data volume can become much larger than has been considered in this paper, to the size of hundreds of terabytes. While our compression-based Marchenko method could assist with using this data volume size, it is advisable to consider only a subset of the data in this case. 

As an additional consequence to the large datasize in 3D, the computational costs of retrieving virtual responses using the Marchenko method increase significantly. While our approach can assist with reducing the time reading data from and writing data to disk, the actual computational process is not affected. Possible solutions are the efficient use of distributed computing \citep{Ravasi2021framework} or the use of GPUs \citep{koehne2021multigpu}. While the use of GPUs for efficient computation has gained much attraction in recent years and is an exciting development, the storage space on GPUs is limited. In the case of large datasets, such as the ones required for the Marchenko method, a substantial number of GPUs would be required. Furthermore, the transfer of data to and from the GPUs can be a significant bottleneck. Further technological developments of GPUs could make this approach more viable in the future.
Current state-of-the-art GPUs would require efficient computation distribution to minimize idle times or an effective compression algorithm could be developed that would allow for the application of the Marchenko method without decompression.

One of the other approaches for computational efficiency that we presented in this paper is the use of virtual plane-wave sources instead of virtual point sources. It should be noted that this approach is a most effective alternative for retrieving the response of a large number of virtual source points in the subsurface that are on the same depth level. On the other hand, when this number is sufficiently small, using the virtual point sources yields a higher accuracy and more angle information. Datasets recorded by dense OBN arrays could benefit greatly from the plane-wave approach, however, even for more traditional streamer setups, the approach can yield benefits, especially for large 3D apertures.

The imaging strategy that we have applied in this paper is based on retrieving the local reflectivity, either at a point or a depth level, and using this to construct the structural image. An alternative to this approach would be to redatum the sources and receivers to a single depth level using the MDD approach, so that a reflection response of the target zone without overburden contributions is constructed. A more straightforward imaging technique such as Reverse Time Migration can then be used to image the target zone. This approach works best if the strongest reflectors are located in the overburden and the target zone is well defined. The area below the overburden can still cause internal multiples however, so care needs to be taken when this approach is employed.

Our data are compressed through the use of the ZFP algorithm, which offers accuracy based on tolerance. We used a value of 1e-7, which means that we allow errors in the compression that are 7 orders of magnitude lower than the maximum amplitude. As we are dealing with single-point precision, this gives us the maximum accuracy possible. By setting the tolerance to a higher value, the data can be compressed further, however, this will introduce numerical errors into the compressed data. Depending on the reduction of data space and the desired accuracy, this may be acceptable for certain applications, however, this is beyond the scope of this paper.

\section*{Conclusions}
We have discussed the implementation of the 3D Marchenko method while taking into account several practical considerations. The standard Marchenko implementation requires high quality reflection data and while it is possible to achieve this, the size of the data volume in 3D is considerably large. In order to handle the large data volume and reduce the computational costs of the 3D Marchenko method, we used a ZFP compression algorithm. This reduced the size of the data and limited the loading time. The 3D Marchenko method was then used to retrieve Green's functions in the subsurface. The result closely matched a reference result and as such the retrieved Green's functions could be used for the purpose of imaging. While the result of the imaging shows attenuation of artifacts related to internal multiples, the computational costs of imaging a large section of the subsurface in 3D was high.

To limit the computational costs of 3D Marchenko imaging further, virtual plane-waves were utilized. This approach allows for the imaging of an entire depth level in one go, instead of just a single point in the subsurface. The computational benefit in 2D is already significant, however, in 3D it is even more potent. The retrieval of the plane-waves in 3D has the limitation that only a certain angle of reflectivity can be retrieved, while the standard Marchenko imaging using point sources theoretically retrieves all possible angles that are present in the data. By combining the images associated with different angles of reflection together, the quality of the plane-wave imaging was improved, while still functioning at a significantly lower computational cost than the standard Marchenko imaging. From an image constructed using the plane-wave Marchenko approach a target zone can be efficiently determined and the standard Marchenko approach using virtual point sources can then be employed to obtain an accurate image in this target zone.

Given the size of the data volume and availability of computational resources to process that data, the (target oriented) point source Marchenko method gives the most accurate results. However, if the computing power is limited, the advantage of utilizing efficient approaches to implement the method becomes evident. The use of compressed data and Eikonal solvers can drastically reduce the storage volume that is required for the data, while the plane-wave approach can be used to efficiently determine a target zone, which can then be further explored using standard Marchenko imaging. For a wider application of the Marchenko method, future research should include consideration of how the solutions to the Marchenko method can be efficiently computed, either through practical or theoretical improvements.

\newpage
\bibliographystyle{apalike}
\bibliography{GPY}

\newpage
\appendix

\section{Input for marchenko3D and auxiliary programs} \label{appinp}
\subsection{marchenko3D} \label{appmar}

The \path{marchenko3D} program has the following parameters and options:
 
\footnotesize 
\begin{verbatim}
 MARCHENKO3D - Iterative Green's function and focusing functions retrieval in 3D
 
 marchenko3D file_tinv= file_shot= [optional parameters]
 
 Required parameters: 
 
   First arrival input options:
   file_tinv= ............... direct arrival from focal point: G_d
   file_ray= ................ direct arrival from raytimes
   Shot data input options:
   file_shot= ............... Reflection response (time data): R(t)
   file_shotw= .............. Reflection response (frequency data): R(w)
   file_shotzfp= ............ Reflection response (frequency compressed data): zfp[R(w)]
 
 Optional parameters: 
 
 INTEGRATION 
   ampest=0 ................. Estimate a scalar amplitude correction with depth (=1)
   tap=0 .................... lateral taper focusing(1), shot(2) or both(3)
   ntap=0 ................... number of taper points at boundaries
   fmin=0 ................... minimum frequency in the Fourier transform
   fmax=70 .................. maximum frequency in the Fourier transform
 MARCHENKO ITERATIONS 
   niter=10 ................. number of iterations
 MUTE-WINDOW 
   file_amp= ................ amplitudes for the raytime estimation
   file_wav= ................ Wavelet applied to the raytime data
   above=0 .................. mute above(1), around(0) or below(-1) the travel times of the first arrival
   shift=12 ................. number of points above(positive) / below(negative) travel time for mute
   hw=8 ..................... window in time samples to look for maximum in next trace
   smooth=5 ................. number of points to smooth mute with cosine window
 MUTE-WINDOW 
   plane_wave=0 ............. enable plane-wave illumination function
   src_anglex=0 ............. angle of the plane wave in the x-direction
   src_angley=0 ............. angle of the plane wave in the y-direction
   src_velox=0 .............. velocity of the plane wave in the x-direction
   src_veloy=0 .............. velocity of the plane wave in the y-direction
 REFLECTION RESPONSE CORRECTION 
   scale=2 .................. scale factor of R for summation of Ni with G_d (only for time shot data)
   pad=0 .................... amount of samples to pad the reflection series
 HOMOGENEOUS GREEN'S FUNCTION RETRIEVAL OPTIONS 
   file_homg= ............... output file with homogeneous Green's function 
   The homogeneous Green's function is computed if a filename is given
   file_inp= ................ Input source function for the retrieval
   scheme=0 ................. Scheme for the retrieval
   .......................... scheme=0 Marchenko homogeneous Green's function retrieval with G source
   .......................... scheme=1 Marchenko homogeneous Green's function retrieval with f2 source
   .......................... scheme=2 Marchenko Green's function retrieval with source depending on virtual
                              receiver location
   .......................... scheme=3 Marchenko Green's function retrieval with G source
   .......................... scheme=4 Marchenko Green's function retrieval with f2 source
   .......................... scheme=5 Classical homogeneous Green's function retrieval
   .......................... scheme=6 Marchenko homogeneous Green's function retrieval with multiple G sources
   .......................... scheme=7 Marchenko Green's function retrieval with multiple G sources
   .......................... scheme=8 f1+ redatuming
   .......................... scheme=9 f1- redatuming
   .......................... scheme=10 2i IM(f1) redatuming
   cp=1000.0 ................ Velocity of upper layer for certain operations
   rho=1000.0 ............... Density of upper layer for certain operations
 IMAGING
   file_imag= ............... output file with image 
   The image is computed if a filename is given
 OUTPUT DEFINITION 
   file_green= .............. output file with full Green function(s)
   file_gplus= .............. output file with G+ 
   file_gmin= ............... output file with G- 
   file_f1plus= ............. output file with f1+ 
   file_f1min= .............. output file with f1- 
   file_f2= ................. output file with f2 
   file_ampscl= ............. output file with estimated amplitudes 
   file_iter= ............... output file with -Ni(-t) for each iteration
   compact=0 ................ Write out homg and imag in compact format
   .......................... WARNING! This write-out cannot be displayed with SU
   zfp=0 .................... Write out the standard output in compressed zfp format
   tolerance=1e-7 ........... accuracy of the zfp compression,
   verbose=0 ................ silent option; >0 displays info
\end{verbatim}
\normalsize

The input of the 3D Marchenko method is similar to the 2D implementation, requiring reflection data and the first arrival time from the focal point. However, due to the large size of 3D data, loading the reflection data from disk using the \path{file_shot} option can become time-consuming. To mitigate this problem, the 3D implementation gives two alternate options of loading the data from disk. The first is loading the shot data in the frequency domain using \path{file_shotw}. These data have been pre-transformed to the frequency domain, which avoids the Fourier transform that was required on the shot data in the time domain. Alternatively, the frequency domain data can be compressed using the ZFP algorithm \citep{Lindstrom2014} before they are loaded from disk, to reduce the file size. The code requires one of these three data types as input, and the latter two options can be obtained using the \path{TWtransform} module.

The second required input, the first arrival time, can be passed to the code in two ways. The first is by loading a shot record using the \path{file_tinv} option. An alternative is using the arrival times that are calculated by a 3D Eikonal solver, an example of which is the \path{raytime3D} program that is part of the software distribution, based on the work by \citet{Vidale1990}. The \path{raytime3D} program also computes a geometric spreading factor that can be used to estimate the amplitude of the first arrivals. This file can be read in using the \path{file_amp} option. To approximate seismic broadband data, a wavelet can be read into the code as well using \path{file_wav}. The file needs to contain a wavelet with no time shift and have the same temporal sampling as the reflection data. For both types of input for the first arrival, multiple focal points can be read in at the same time.

The number of iterations required for convergence depends on the reflection\linebreak strengths and on the number of events in the model; a complex model will need more iterations. Typically the number of iterations is chosen between 8 and 20. Setting the \path{verbose=2} option will compute the convergence of the algorithm by printing the energy of the iteration update $N_i(t)$ relative to the initial value $N_0(t)$. The energy in the update term $N_i(t)$ should become smaller in each iteration. 

The \path{marchenko3D} program is capable of computing an image for the focal points and outputting them directly through use of the \path{imaging3D} module of the code. To do this, one can simply set the \path{file_imag} option and the code computes an image point for each given focal-point. 

The program also contains an additional module called \path{homogeneous3D}, which is used for the purpose of retrieving the wavefield between two focal points in the subsurface. By setting the option \path{file_homg} to a correct path, the wavefield is computed according to the scheme set by the option \path{scheme}. Most of these schemes are explained in \citet{brackenhoff2019} and \citet{brackenhoff2019virtual}. All of the schemes require that the module is given input data by the \path{file_inp} option. This input file needs to be sampled at the same positions as the data that are computed in the main \path{marchenko3D} program with the same sample length and distance. It is recommended to run the module for one focal position first before using it on a large amount of focal positions.

To compensate for the transmission losses, an approximate amplitude correction can be used  \citep{Neut2018single}. This estimation can be added to the results by setting the \path{ampest} option equal to 1. By using the \path{file_ampscl} option, the estimated amplitude corrections are written out for each focal point.

The code to produce examples of the Marchenko method on a flat layered model in this paper can be found in the directory \path{marchenko3D/demo/marchenko3D/oneD}. The \path{README} file in that directory explains in detail how to run the scripts. The SEG/EAGE Overthrust model by \citet{aminzadeh1997} can be found on the SEG wiki at \url{https://wiki.seg.org/wiki/SEG/EAGE_Salt_and_Overthrust_Models}.

\subsection{TWtransform} \label{apptw}

\footnotesize{\begin{verbatim}
 TWtransform - Transform data from uncompressed time domain to compressed frequency 
               domain
 
 TWtransform file_in= file_out= [optional parameters]
 
 Required parameters: 
 
   file_in= ................. File containing the uncompressed time domain data
   file_out= ................ Output for the (compressed) frequency domain data
 
 Optional parameters: 
 
   verbose=1 ................ silent option; >0 displays info
   fmin=0 ................... minimum frequency in the output
   fmax=70 .................. maximum frequency in the output
   mode=1 ................... sign of the frequency transform
   zfp=0 .................... (=1) compress the transformed data using zfp
   tolerance=1e-3 ........... accuracy of the zfp compression, smaller values 
                              give more accuracy to the compressed data but 
                              will decrease the compression rate
   weight=2.0 ............... scaling of the reflection data
\end{verbatim}}
\normalsize

The \path{TWtransform} program is intended to reduce the file size of the reflection data and to limit the amount of time that is required for loading the reflection data. The program transforms the time data to the frequency domain, which already reduces the file size depending on the frequency range set by \path{fmin} and \path{fmax}. The file size can be further reduced by using the \path{zfp} option, which will compress the frequency data using the ZFP compression by \citet{Lindstrom2014}, with an accuracy set by the input parameter \path{tolerance}. The program also puts a custom header on the data to further reduce the file size.

\end{document}